\documentclass[a4paper,usenatbib]{mnras}

\usepackage[T1]{fontenc}
\usepackage{ae,aecompl}

\usepackage[dvipsnames]{xcolor}

\usepackage{graphicx}	
\usepackage{amsmath}	
\usepackage{amssymb}	
\usepackage{BibDef}
\usepackage{ragged2e}
\usepackage{booktabs}
\usepackage{multirow}
\usepackage{mathrsfs}
\usepackage{comment}
\usepackage{ulem}

\usepackage{enumitem}

\graphicspath{{figures/}} 
\def\msun{\!{\rm M}_\odot}
\def\lsim{\mathrel{\rlap{\lower 3pt \hbox{$\sim$}} \raise 2.0pt \hbox{$<$}}}
\def\gsim{\mathrel{\rlap{\lower 3pt \hbox{$\sim$}} \raise 2.0pt \hbox{$>$}}}

\defcitealias{Venemans_et_al_2020}{V20}

\def\noAGN{\texttt{noAGN}}
\def\AGNcone{\texttt{AGNcone}}

\title[Enhanced star formation in $z\sim 6$ quasar companions]{Enhanced star formation in $z\sim 6$ quasar companions}

\author[T. Zana et al.]{Tommaso Zana$^{1}$\thanks{\href{mailto:tommaso.zana@sns.it}{tommaso.zana@sns.it}}, Simona Gallerani$^{1}$, Stefano Carniani$^{1}$, Fabio Vito$^{1}$, 
\newauthor Andrea Ferrara$^{1}$, Alessandro Lupi$^{2,3}$, Fabio Di Mascia$^{1}$, Paramita Barai$^{4,5}$ \\
$^{1}$Scuola Normale Superiore, Piazza dei Cavalieri 7, I-56126 Pisa, Italy\\
$^{2}$Dipartimento di Fisica G. Occhialini, Universit\`{a} di Milano-Bicocca, Piazza della Scienza 3, IT-20126 Milano, Italy\\
$^{3}$INFN, Sezione di Milano-Bicocca, Piazza della Scienza 3, IT-20126 Milano, Italy\\
$^{4}$Centro de Ci\^{e}ncias Naturais e Humanas - Universidade Federal do ABC, Av. dos Estados 5001, Santo Andr\'{e} - SP, 09210-580, Brazil\\
$^{5}$N\'{u}cleo de Astrof\'{i}sica - Universidade Cidade de S\~{a}o Paulo, Rua Galv\~{a}o Bueno 868, S\~{a}o Paulo - SP, 01506-000, Brazil\\
}
\date{Accepted XXX. Received YYY; in original form ZZZ}

\pubyear{2022}

\begin{document}
\label{firstpage}
\pagerange{\pageref{firstpage}--\pageref{lastpage}}
\maketitle

\begin{abstract}
Quasars powered by supermassive black holes (MBH, $>10^8~\msun$) at $z\sim 6$ are predicted to reside in cosmic over-dense regions. However, observations so far could not confirm this expectation due to limited statistics. The picture is further complicated by the possible effects of quasar outflows (i.e. feedback) that could either suppress or stimulate the star formation rate (SFR) of companion galaxies, thus modifying the expected bias. Here we quantify feedback effects on the properties and detectability of companions by comparing cosmological zoom-in simulations of a quasar in which feedback is either included or turned-off. 
With respect to the no-feedback case, companions (a) directly impacted by the outflow have their SFR increased by a factor $2-3$, and (b) tend to be more massive.
Both effects shift the [CII]158$\mu$m and UV luminosity functions toward brighter magnitudes. 
This leads us to conclude that quasar feedback slightly increases the effective quasar bias, boosting the number density of observable quasar companions, in agreement with what has been found around the brightest quasars of recent ALMA [CII] surveys. Deeper observations performed with JWST and/or ALMA will improve the statistical significance of this result by detecting a larger number of fainter quasar companions.
\end{abstract}


\begin{keywords}
galaxies: evolution -- galaxies: high-redshift -- galaxies: star formation -- quasars: general -- quasars: supermassive black holes -- methods: numerical
\end{keywords}


\section{Introduction}


Massive black holes (MBHs) are thought to be ubiquitous in the centre of massive galaxies at all redshifts \citep[][and references therein]{Kormendy_Ho_2013}.
Through gas accretion, they are the engines responsible for the huge emission of galactic centres as active galactic nuclei (AGN).
Extremely bright quasars are AGN with enormous luminosities ($L_{\rm bol} > 10^{46}$~erg~s$^{-1}$), likely powered by the most massive MBHs, with masses $M_{\rm BH}$ ranging from $10^{8}$ to $10^{10}~\msun$.

The number of observed bright quasars at $z>6$ has substantially grown in the last years \citep[e.g,][]{Wu_et_al_2015, Venemans_et_al_2015, Jiang_et_al_2015, Wu_et_al_2015, Carnall_et_al_2015, Reed_et_al_2015, Matsuoka_et_al_2016, Reed_et_al_2017, Banados_et_al_2018}, proving that MBHs with $M_{\rm BH} \sim 10^{9}~\msun$ were already in place when the Universe was less than $950$~Myr old.

The fast growth of these compact objects represents an exciting and fully open problem in modern astrophysics and various models have been proposed to solve the issue.
For a long time, it was assumed that MBH seeds were provided by ($i$) the massive remnants of Pop III stars \citep[see, e.g.][]{Madau_Rees_2001, Haiman_Loeb_2001, Heger_et_al_2003, Volonteri_et_al_2003, Madau_et_al_2004}, formed at $z\gsim20$.
However this scenario is problematic, since the accretion would have to either continue unperturbed at the Eddington rate, maintaining a very low radiative efficiency \citep[not to  excessively hinder the feeding process; see, e.g.][]{Tanaka_Haiman_2009}, or ($ii$) exceed the Eddington limit \citep[see, e.g.][]{Madau_et_al_2014, Volonteri_et_al_2015, Lupi_et_al_2016, Pezzulli_et_al_2017}.
Other models predict the ($iii$) existence of very massive MBH seeds ($M_{\rm BH} \gsim 10^{4}~\msun$, possibly originated from direct collapse; see e.g. \citealt{Loeb_Rasio_1994, Bromm_Loeb_2003, Koushiappas_et_al_2004, Spaans_Silk_2006, Mayer_et_al_2010, Mayer_et_al_2015}). 

The steady and efficient gas inflow, necessary for the fast MBH growth, would be strongly favoured within highly massive dark matter (DM) halos \citep[see][]{Efstathiou_Rees_1988}.
Accordingly, theoretical models including numerical simulations with MBH accretion and feedback prescriptions \citep[see e.g.][]{Barkana_Loeb_2001, Springel_et_al_2005a, Sijacki_et_al_2009, Di_Matteo_et_al_2012, Costa_et_al_2014, Weinberger_et_al_2018, Ni_et_al_2018, Marshall_et_al_2019, Ni_et_al_2020} predict high-$z$ MBHs to reside in the most massive DM-halos with $M_{\rm vir} > 10^{12}-10^{13}~\msun$, corresponding to fluctuations above $3-4\sigma$ in the cosmic density field \citep{Barkana_Loeb_2001}.
A connection between MBHs and their host DM halo has been studied also via clustering
measurements of AGN. Various large surveys, from the optical to the X-ray band, have shown that AGN reside in massive halos with masses $M_{\rm vir} = 10^{12}-10^{13}~\msun$ at low ($z\sim0.1$) and higher (up to $z\approx2$) redshift \citep[e.g.][]{Hickox_et_al_2009, Ross_et_al_2009, Cappelluti_et_al_2010, Allevato_et_al_2011, Allevato_et_al_2012}.
As a consequence, quasars at high redshift should be part of large scale structures marked by a significant over-density in galaxy number count, that may extend over Mpc scales \citep[e.g, ][]{Costa_et_al_2014}. However, this expectation has not been conclusively confirmed by observations.

On the one hand, various observations corroborated the high-density scenario hypothesis (see, e.g. \citealt{Steidel_et_al_2005, Capak_et_al_2011, Swinbank_et_al_2012, Husband_et_al_2013} for $2<z<5$ and \citealt{Kim_et_al_2009, Morselli_et_al_2014, Balmaverde_et_al_2017, Decarli_et_al_2017, Decarli_et_al_2018, Mignoli_et_al_2020, Venemans_et_al_2020}, for $z \sim 6 $ and above).
In particular, \citet{Venemans_et_al_2020} followed up with the Atacama Large Millimeter Array (ALMA) a sample of 27 $z\sim 6$ quasars, previously detected in [CII], discovering 17 [CII] bright galaxies and finding that some of the quasars present multiple (2-3) companions.
Furthermore, whereas no dual AGN has been confirmed so far at $z>6$ \citep[][]{Connor_et_al_2019, Connor_et_al_2020, Vito_et_al_2019a, Vito_et_al_2021}, several of them have been detected up to ${z \sim 5}$ (see \citealt{Koss_et_al_2012, Vignali_et_al_2018, Silverman_et_al_2020}).

On the other hand, more than a few observations did not reveal significant galaxy count over-densities around quasars (see, e.g., \citealt{Francis_Bland-Hawthorn_2004} and \citealt{Simpson_et_al_2014a} at $z\sim 2$, \citealt{Uchiyama_et_al_2018} at $z=4$, \citealt{Kashikawa_et_al_2007} and \citealt{Kikuta_et_al_2017} at $z\sim 5$, \citealt{Banados_et_al_2013} and \citealt{Mazzucchelli_et_al_2017b} at $z>5.7$)
and some theoretical studies consistently suggested that MBHs does not have to inevitably lie in the most massive halos \citep[see, for instance][]{Fanidakis_et_al_2013, Orsi_et_al_2016, Di_Matteo_et_al_2017, Habouzit_et_al_2019}.

These conflicting observational results could be explained by the difficulty in detecting the satellites in the neighborhood of a luminous quasar, where the strong AGN feedback can play a fundamental role in shaping their baryonic component.
AGN feedback could affect the MBH environment well beyond the host galaxy scale radius and significantly modify the star formation (SF) activity of the orbiting companions \citep[see][]{Martin-Navarro_et_al_2019, Martin-Navarro_et_al_2021}.
\citet{Dashyan_et_al_2019} explicitly investigated, in cosmological hydrodynamical simulations, the AGN-driven quenching effect within galactic satellites at low redshift ($z<3$).
The authors claim that AGN winds can decrease the SF process in AGN companions, by sweeping away their gas, out to five times the virial radius of the central galaxy.
Conversely, \citet{Gilli_et_al_2019} observed a radio AGN at $z=1.7$, finding a possible positive effect of its feedback on the star formation of its companion, over a scale $\gsim450$~kpc. Furthermore, \citet{Fragile_et_al_2017} performed an isolated simulation in order to explain the positive feedback observed by \citet{Croft_et_al_2006} in a star-forming galaxy, where the star formation is triggered by the radio-jets of the nearby NGC--541. 

At higher redshift, the direct effect of AGN feedback on the satellite galaxies remains unclear.
Several studies found, both through observations \citep[][]{Kashikawa_et_al_2007}, and numerically \citep[][]{Efstathiou_1992, Thoul_Weinberg_1996, Okamoto_et_al_2008}, that the ionizing background produced by the quasar is able to inhibit SF in satellites, or even suppress their assembling.
In the most extreme cases, satellite halos would still populate the quasar environment, without being detectable because of their low gas and stellar content, especially given the limited sensitivity of modern instruments.
In this context, the upcoming James Webb Space Telescope (JWST, \citealt{Gardner_et_al_2006}) will allow us to observe fainter AGN companions, hopefully alleviating our sensitivity bias. With a primary mirror of about 6.5 meters, JWST reaches a sensitivity 3-5 times higher than that of Hubble Space Telescope at 1$\mu$m, enabling the detection of the rest-frame UV emission that arises from distant faint star-forming galaxies with relative short exposure time.

Numerical simulations are powerful tools to study the AGN effect on the surrounding galaxies, for they provide the complete spatial and time distribution of matter, as well as fundamental self-consistent sub-grid models, such as SF, stellar feedback, cooling processes, etc.
Recently, enormous steps forward have been made in this field and several works successfully portrayed the formation and evolution of bright quasar hosts at high redshift.
For instance, \citet{Di_Matteo_et_al_2017} investigated the accretion efficiency of high-$z$ MBHs, by employing advanced refinement techniques, focussing on the effects of AGN feedback on the host galaxy. 
\citet{Curtis_Sijacki_2016} and \citet{Van_der_Vlugt_Costa_2019} studied how AGN feedback shapes the dynamical components of the galaxy.
\citet{Costa_et_al_2014} and \citet{Smidt_et_al_2018} analysed how AGN X-ray luminosity and feedback affect negatively SF in the host system.
\citet{Lupi_et_al_2019} and \citet{Lupi_et_al_2021} detailed the consequences of AGN thermal feedback on the host interstellar medium (ISM).
Differently, \citet{Richardson_et_al_2016}, \citet{Barai_et_al_2018}, and \citet{Valentini_et_al_2021} studied the effect of AGN feedback on the formation and evolution of a proto-cluster, using either different numerical methods or different feedback prescriptions.
Finally, \citet{Costa_et_al_2020} developed a state of the art model to describe AGN-driven small-scale winds and tested it in isolated simulations.

In this work, we analyse the environment of a powerful quasar at $z\gsim6$, resulting from the simulations by \citet{Barai_et_al_2018}.
To assess the possible effect of the AGN feedback on the surrounding companion galaxies, we take advantage of the \citet{Barai_et_al_2018} suite and compare the quasar environment with a control run in which MBHs are not seeded. In detail, our goal is to ($i$) evaluate the impact of quasar feedback on its environment, far beyond its host galaxy radius, and ($ii$) provide theoretical predictions on the expected UV and rest-frame FIR luminosities of the neighbour satellites.

This paper is organized as follows: in Section~\ref{hydro_sim}, we describe the numerical model adopted in this work and introduce the runs with and without AGN, hereafter called \AGNcone{} and \noAGN{}, respectively;
in Section~\ref{sec:sample} we present the sample of satellites and statistically analyse their redshift evolution and how their properties (e.g. number of satellites, star formation rate, stellar mass, gas mass, metallicity) are affected by their position in the proto-cluster;
in Section~\ref{sec:effect_individual}, we focus on the effect of AGN feedback on individual satellites, and present an interpretation of our results in Section~\ref{sec:discussion};
in Section~\ref{sec:observations} we discuss the observational properties of the galaxy group;
finally, we summarise our findings and draw our conclusions in Section~\ref{sec:conclusions}.


\section{Hydrodynamic simulations}
\label{hydro_sim}

The two runs analysed in this work belong to a suite of zoom-in cosmological simulations \citep[][]{Barai_et_al_2018} built to follow the formation of a massive galaxy proto-cluster at $z\simeq6$ through the smoothed particle hydrodynamics $N$-body code {\textsc{gadget-3}} \citep{Springel_2005, Springel_et_al_2008}.

Both the simulations share the same initial conditions, generated through the code \textsc{music}
\citep[see][]{Hahn_Abel_2011} and assume the same recipe for the sub-grid physics, with the exception of the MBH prescription.
In particular, the cosmological parameter set refers to a flat $\Lambda$CDM Universe with $\Omega_{\rm M,0} = 0.3089$, $\Omega_{\Lambda,0} = 1-\Omega_{\rm M,0} = 0.6911$, $\Omega_{\rm b,0} = 0.0486$, and $H_{0} = 67.74$~Mpc s$^{-1}$ \citep[][results XIII]{Plank_2015}.
The simulated box of side $500$~comoving Mpc has been evolved with only DM particles from $z=100$, till $z\lesssim6$, with an initial mass resolution of $2\times10^{10}~\msun$ per particle and a softening length of $48.72$ comoving kpc.
There, the Lagrangian volume of the most massive proto-cluster has been identified (through a \textit{Friends-of-Friends} algorithm) with a virial mass $M_{\rm vir} \simeq 10^{12}\, \msun$ and a comoving virial radius $r_{\rm vir}\simeq 511$~kpc, traced back to $z=100$, refined and re-simulated along with baryons.
In the most refined region -- set to be, originally, a cube of side $5.21$~Mpc --, the mass resolution is given by $m_{\rm DM} = 7.54 \times 10^6~\msun$ for 591408 DM particles, and $m_{\rm gas} = 1.41 \times 10^6~\msun$ for the same number of gas particles, whereas the spatial resolution is set by the gravitational softening length of all particle species ($\epsilon\simeq 1.476$ comoving kpc).
The adaptive smoothing length is computed according to the standard prescription by \citet{Springel_et_al_2008} and its minimum value is set to $0.001\epsilon$.

The whole suite implements radiative heating and cooling using the rates provided in the tables of \citet{Wiersma_et_al_2009} in ionization equilibrium. 
Metal-line cooling is also considered.
Eleven element abundances (H, He, C, Ca, O, N, Ne, Mg, S, Si, Fe) are followed according to the receipt of \citet{Tornatore_et_al_2007} in the presence of a redshift-dependent cosmic ionizing background \citep{Haardt_Madau_2012}.

SF is modelled following the multiphase recipes by \citet{Springel_Hernquist_2003}. More specifically, gas particles denser than $n_{\rm SF} = 0.13$~cm$^{-3}$ are converted into collisionless star particles according to the stochastic scheme of \citet{Katz_et_al_1996}.
Each spawned star particle represents a stellar population described by a \citet{Chabrier_2003} initial mass function in the mass range $0.1-100~\msun$.
Stars are allowed to explode as supernovae (SN) releasing kinetic energy via a constant-velocity outflow with $v_{\rm SN} = 350$~km~s$^{-1}$ \citep[see][]{Barai_et_al_2015, Biffi_et_al_2016}.
Metal enrichment of the interstellar medium is provided by Type Ia SN ($0.8<M /\msun<8$) considering a fraction of binary of $1/10$, according to \citet{Thielemann_et_al_2003}, Type II SN ($M /\msun>8$) according to \citet{Woosley_Weaver_1995}, and winds from asymptotic giant branch stars following \citet{van_den_Hoek_Groenewegen_1997}.

The two simulations differ only in terms of the prescription used to describe massive black holes.
In detail, in \noAGN{} only cooling, metal enrichment, star-formation, and SN feedback are included, with no prescription for MBHs.

By contrast, in the \AGNcone{} run MBHs are represented as sink particles that can form and grow both via accretion of gas and through mergers with other MBHs.
In particular, a MBH is seeded in a halo when: ($i$) the halo does not host any other MBH, ($ii$) the halo virial mass is $M_{\rm h} \geq 10^{9}~\msun$ (i.e. the halo is properly resolved).
When these conditions are satisfied, a $M_{\rm BH} = 10^{5}$~$\!\msun$ MBH is placed at the centre of mass of the halo.
The simulation implements a \textit{repositioning} algorithm as in \citet{Springel_et_al_2005b, Schaye_et_al_2015}.

Every MBH can accrete mass from the surrounding medium via the classical Bondi--Hoyle--Lyttleton accretion rate \citep{Hoyle_Lyttleton_1939, Bondi_Hoyle_1944, Bondi_1952}:
\begin{equation}
    \dot{M}_{\rm Bondi} = \frac{4 \pi G^2 M^2_{\rm BH} \rho}{(c^2_{\rm s}+v^2)^{3/2}}, 
    \label{eq:Mbondi}
\end{equation}
where $G$ is the gravitational constant, $\rho$ is the gas density, $c_{\rm s}$ the sound speed, and $v$ the gas relative velocity with respect to the MBH\footnote{In the code, gas particles are swallowed according to a stochastic method described in \citet{Springel_et_al_2005b}.}.
The accretion rate is multiplied by a boost factor of $100$, analogously to what has been done, e.g., in \citet{Springel_et_al_2005b} and it is capped to the Eddington limit,
\begin{equation}
    \dot{M}_{\rm Edd} = \frac{4 \pi G M_{\rm BH} m_{\rm p}}{\epsilon_{r} \sigma_{\rm T} c},
    \label{eq:Medd}
\end{equation}
where $m_{\rm p}$ is the proton mass, $\sigma_{\rm T}$ the Thomson cross section, and $c$ the speed of light in vacuum.
$\epsilon_{\rm r}$ is the radiative efficiency set equal to $0.1$ 
\citep[see the average efficiency for an optically thick and geometrically thin accretion disks by][]{Shakura_Sunyaev_1973}.
During the accretion process a MBH radiates a fraction $\epsilon_{\rm r}$ of the accreted rest-mass energy 
\begin{equation}
    L_{\rm rad} = \epsilon_{\rm r} \dot{M}_{\rm acc} c^2,
    \label{eq:Lrad}    
\end{equation}
where $\dot{M}_{\rm acc}$ is the rate of the inflowing gas onto the MBH, and a fraction $\epsilon_{\rm f}$ of this luminosity is coupled to the interstellar medium as feedback energy:
\begin{equation}
    \dot{E}_{f} = \epsilon_{\rm f} L_{\rm rad},
    \label{eq:Efeed}    
\end{equation}
where $\epsilon_{\rm f} = 0.05$ as in, e.g., \citet{Di_Matteo_et_al_2008}.
Feedback is kinetic and is modelled through the ejection, in a bi-cone with a half-opening angle of $45$~degrees, of a certain mass of gas $M_{\rm w}$ with a fixed initial velocity of $v_{\rm w} = 10^{4}$~km s$^{-1}$, such that the kinetic luminosity $\frac{1}{2}\dot{M_{\rm w}}v_{\rm w}^2$ is equal to $\dot{E}_{f}$.
The direction of emission is random and it is associated to the MBH when it is seeded.
We note that this assumption is supported by several studies showing little or no alignment  between the outflow/jet axis and the large-scale angular momentum of the host galaxy (see, e.g., \citealt{Hopkins_et_al_2012}, and references therein).

To summarise, in the control run \noAGN{} only SN feedback is included, whereas \AGNcone{} incorporates both SN and AGN feedback. 
Different feedback prescriptions result into different host galaxy properties \citep[see][for a detailed discussion]{Barai_et_al_2018}.
Here, we focus our analysis on the quasar environment.
The properties of the two runs at their last snapshot are outlined in Table~\ref{tab:summary}.

\begin{table*}
\centering
\caption{Summary table for the two analysed runs at $z=6$. From left to right: ($i$) name of the simulation, ($ii$) presence of SN feedback; ($iii$) presence of AGN feedback, ($iv$) total number of MBHs, ($v$) accretion rate of the most accreting MBH, ($vi$) mass of the most massive MBH, ($vii$) stellar mass and ($viii$) SFR of the central dominant galaxy. At $z=6$ in \AGNcone{}, the most massive MBH is also the most accreting one, even if this is not always true at higher redshift.}
\label{tab:summary}
\begin{tabular}{cccccccc}
	\hline	
	Run & SF/SN feedback & AGN feedback & \# MBH & $\dot{M}_{\rm acc} [\msun$ yr$^{-1}]$ & $M_{\rm BH} [\msun]$ & $M_{*}^{\rm cD} [\msun]$ & $SFR^{\rm cD} [\msun$ yr$^{-1}]$\\
	\hline \hline
	\noAGN{} & yes & no & 0 & 0 & 0 & 1.5$\times 10^{11}$ & 664\\
	\hline
	\AGNcone{} & yes & yes & 723 & 57.6 & 4.85$\times 10^9$ & 6.5$\times 10^{10}$ & 116\\
	\hline
\end{tabular}
\end{table*}


\section{The satellite sample}
\label{sec:sample}

\begin{figure}
    \includegraphics[width=0.49\textwidth]{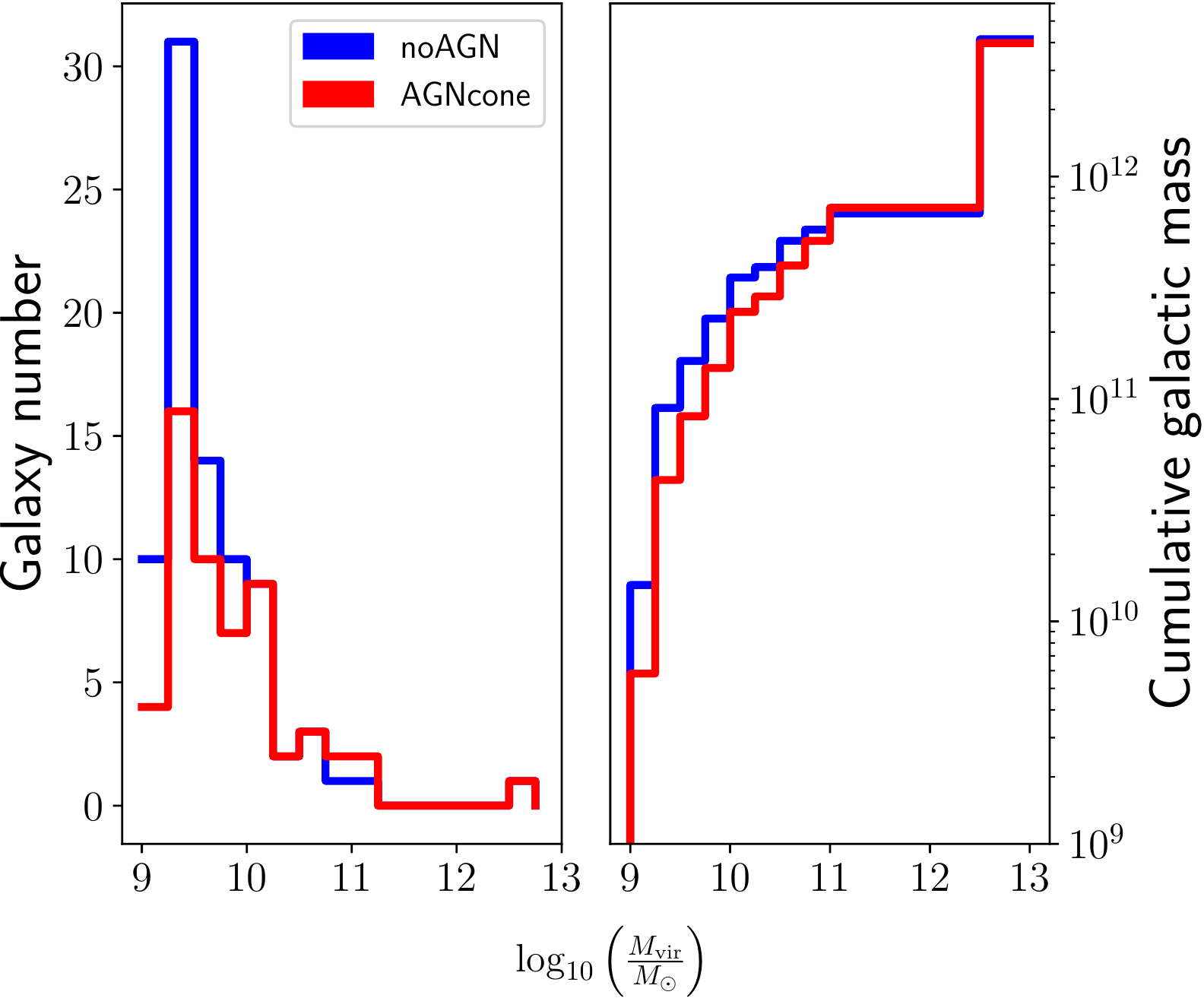}
    \caption{{\it Left panel}: virial mass distribution of the galaxy samples in \noAGN{} (blue) and \AGNcone{} (red) at $z=6$.
    {\it Right panel}: cumulative virial mass function for the same samples.
    Note the highest mass bin, containing the cD galaxy of the proto-cluster.}
    \label{fig:Mvir_PDFs}
\end{figure}

We identify galaxies and their related dark matter halos through the \textsc{AMIGA} halo finder code \citep[see,][]{Knollmann_et_al_2009}, using a minimum of 20 bound particles to define a halo.
The merger tree for each halo at $z\simeq6$ is built by tracing back in time the constituent DM particles: their ID is matched in the progenitor structures in the previous snaphots.
Baryons are assigned to their related galaxy ($i$) when a gas or a stellar particle is found within $\beta r_{\rm vir}$ of a given halo, where $r_{\rm vir}$ is the virial radius of the halo and $\beta=0.3$ (similarly to what has been done in \citealt{Rosdahl_et_al_2018} and in \citealt{Costa_et_al_2019}) and ($ii$) when the velocity of the considered particles is lower than the escape velocity, determined by the DM potential well\footnote{The gravitational potential here is evaluated through the analytical integration of a Navarro-Frank-White profile \citep[][]{Navarro_et_al_1996}.}.

In the following analysis we select only those satellites with a contamination from low-resolution DM\footnote{We cannot exclude the possibility of contamination-induced effects on the galactic satellites that orbits farther out in the refined zone of the simulation volume with respect to the main galaxy. For this reason we impose such constraint.
The number of excluded galaxies is never larger than $1-2$ per snapshot.} particles lower than 20 percent in mass.

We consider only galaxies with $M_{\rm vir} > 10^{9}~\msun$ and with a minimum stellar mass of $M_{*} = 10^{7}~\msun$.
These thresholds are a good compromise to minimize the numerical errors, still maximizing the number of objects for statistical significance.

The selected samples at $z=10$ consist of 35 galaxies for run \noAGN{} and 36 for run \AGNcone{}.
At $z\simeq6$ the difference between the number of satellites in the two runs becomes notable, being in \noAGN{}, 30~percent larger than in \AGNcone{} (82 versus 56).
We detail this difference in \S~\ref{subsec:redshift_evolution_number}.

We find that at $z\simeq6$, the virial radii of satellites range from about 2 to about 21~kpc and masses from $10^{9}$ to $10^{11}~\msun$.
The distribution of the galactic virial masses at $z=6$ is shown in the left panel of Figure~\ref{fig:Mvir_PDFs} for both the analysed runs.
Even though the initial conditions are identical and the virial mass is dominated by the DM component, which is far less sensitive to the feedback prescription than baryons, small but appreciable variations are present.
Feedback from MBHs seems to have a remarkable effect both in redistributing the baryonic component among the companions and in driving several satellites to coalescence, resulting in larger systems.
After almost 1~Gyr from the beginning of the simulation, \AGNcone{} galaxies are less peaked around $\sim 10^{9}~\msun$, than \noAGN{}, whereas the cumulative mass of the galaxy populations (right panel of Figure~\ref{fig:Mvir_PDFs}) shows that the total mass is conserved.
This result demonstrates that no mass is lost in \AGNcone{} trough tidal disruption processes of smaller halos.

At $z\simeq6$, about half of the galaxies in the sample of \AGNcone{} hosts at least a MBH, whose mass ranges from $\sim10^{5}~\msun$, to $4.8\times10^9~\msun$.
There are both quiescent MBHs (or with a negligible accretion rate) and strongly accreting MBHs with $~\sim60-70~\msun$~yr$^{-1}$ (see Table~\ref{tab:summary}). 
In particular, the most accreting MBH does not remain in the same object during the whole evolution of the galactic proto-cluster and it is not always the most massive one.

To summarise, the galaxies examined here exhibit a complex network of AGN, whose emitted energy varies with time and whose geometrical distribution, along with the preferred direction of emission of feedback, requires a proper model to study their influence on the satellite population.

\subsection{Redshift evolution of companions}
\label{subsec:redshift_evolution_number}

\begin{figure}
    \centering
    \includegraphics[width=0.47\textwidth]{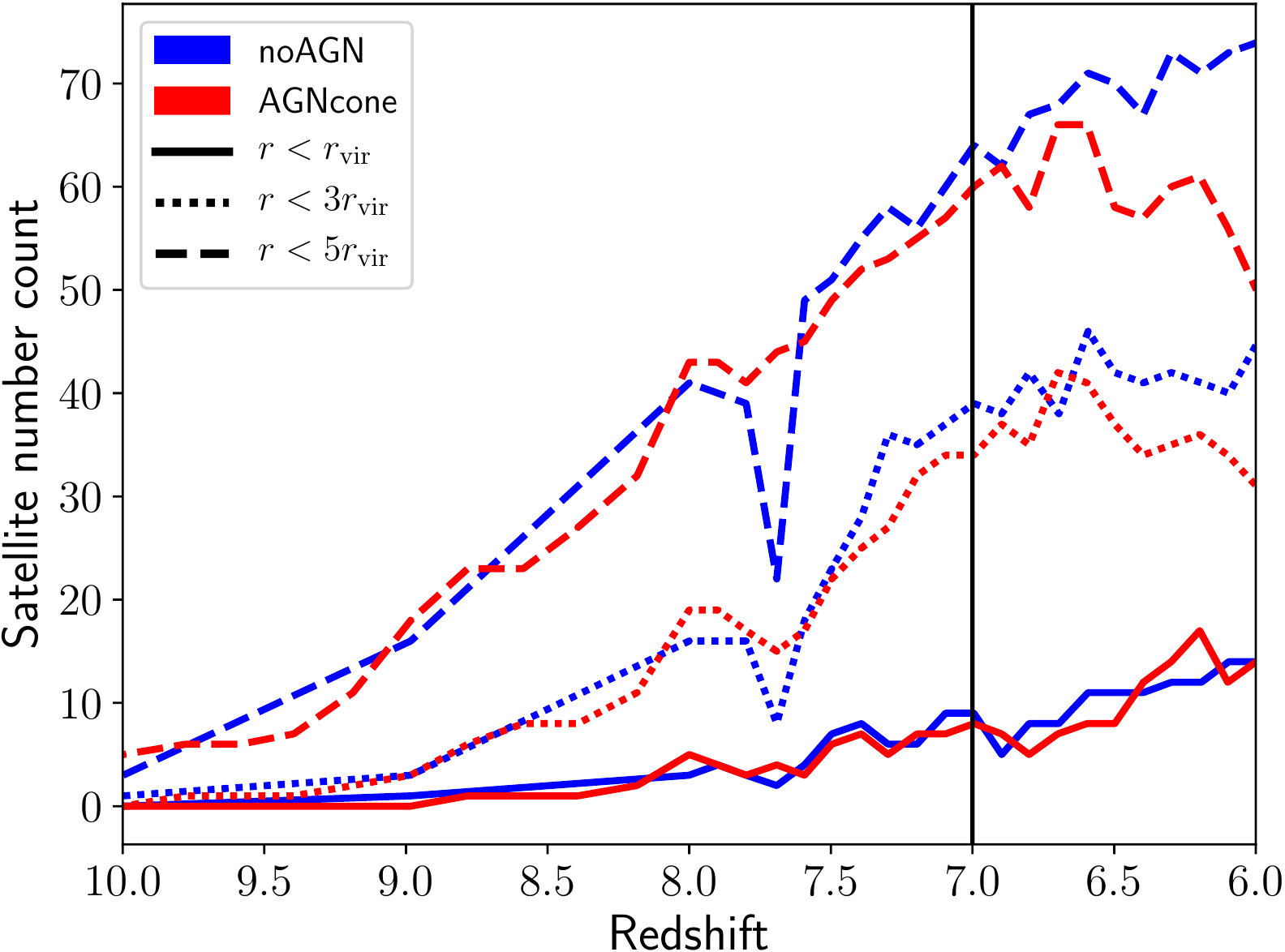}
    \caption{Number counts of satellites as a function of redshift. Solid, dotted, and dashed lines indicate the number count evaluated within 1, 3 and 5 virial radii, respectively.
    Blue lines mark the trend for \noAGN{}, whereas red lines refer to \AGNcone{}.
    Differences between the runs significantly increase after about $z=7$, here highlighted with a vertical black line.}
    \label{fig:gal_count}
\end{figure}

\begin{figure*}
    \centering
    \includegraphics[width=\textwidth]{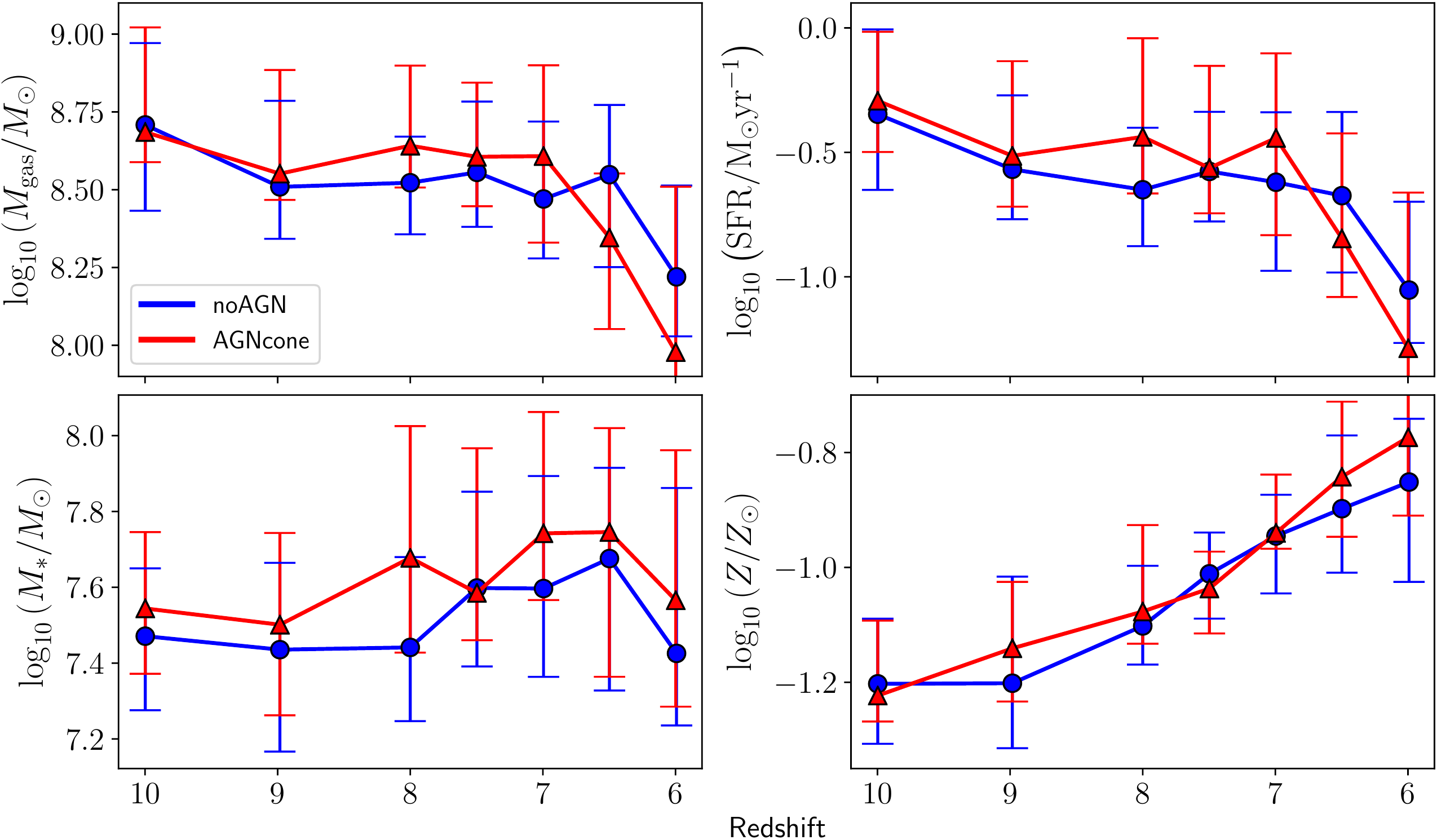}
    \caption{Redshift evolution of satellite median properties (\noAGN{} in blue, \AGNcone{} in red). Clockwise from top left: gas mass, $M_{\rm gas}$, star formation rate, SFR, total gas metallicity, $Z$ in solar units ($Z_{\odot} = 0.0196$, according to \citealt{Vagnozzi_2019}), and stellar mass, $M_{*}$. The error bars represent the $30$-th (error bar-lower end) and the $70$-th percentile (error bar-upper end) of the distribution. We note that the median stellar mass in the \AGNcone{} run has a trend compatible with a higher cumulative SF history with respect to \noAGN.}
    \label{fig:redshift_evolution}
\end{figure*}

\begin{figure*}
    \centering
    \includegraphics[width=\textwidth]{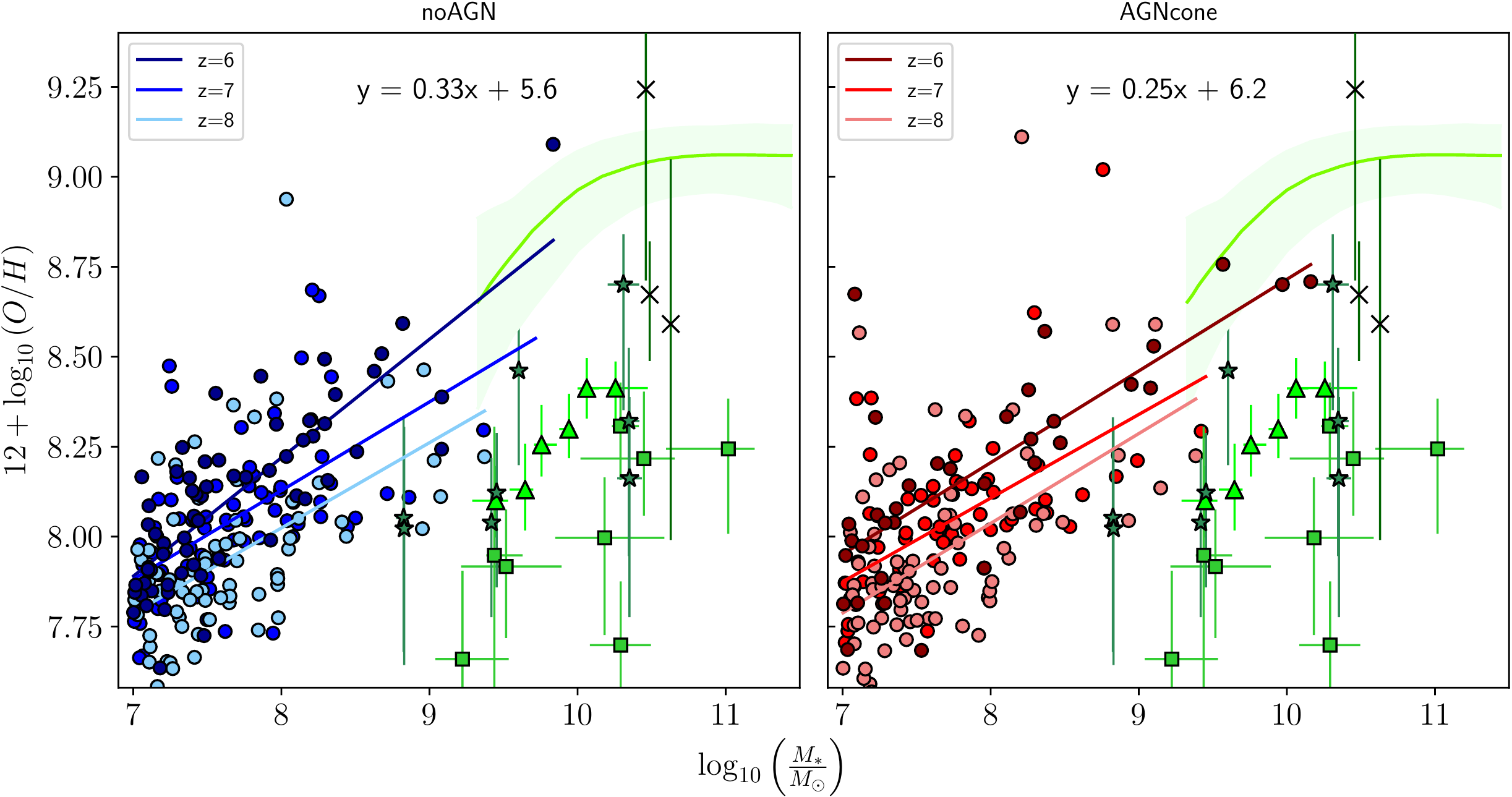}
    \caption{Stellar mass-metallicity relation for \noAGN{} ({\it left panel}; blue, light blue, and cyan points for $z=6, 7, 8$, respectively) and \AGNcone{} ({\it right panel}; red, light red, and pink points for $z=6, 7, 8$, respectively). A linear fit of the same colour is superimposed to each distribution.
    The best-fit equations for $z=6$ populations are $y=0.33x+5.6$ and $y=0.25x+6.2$ for \noAGN{} and \AGNcone{}, respectively.
    For comparison, we plot over each panel some of $M_*-Z$ relations from the gas phase in star-forming galaxies: \citet{Mannucci_et_al_2010} at $z=0$ (light green line with a shaded region representing $90$~percent of the SDSS galaxies), \citet{Cullen_et_al_2014} at $z\gsim2$ (triangles), \citet{Maiolino_et_al_2008} at $z\sim3.5$ (squares), \citet{Faisst_et_al_2016} at $z\sim5$ (stars), and \citet{Harikane_et_al_2020} at $z\sim6$ (crosses).}
    \label{fig:Mstar_Z}
\end{figure*}

\begin{figure*}
    \centering
    \includegraphics[width=\textwidth]{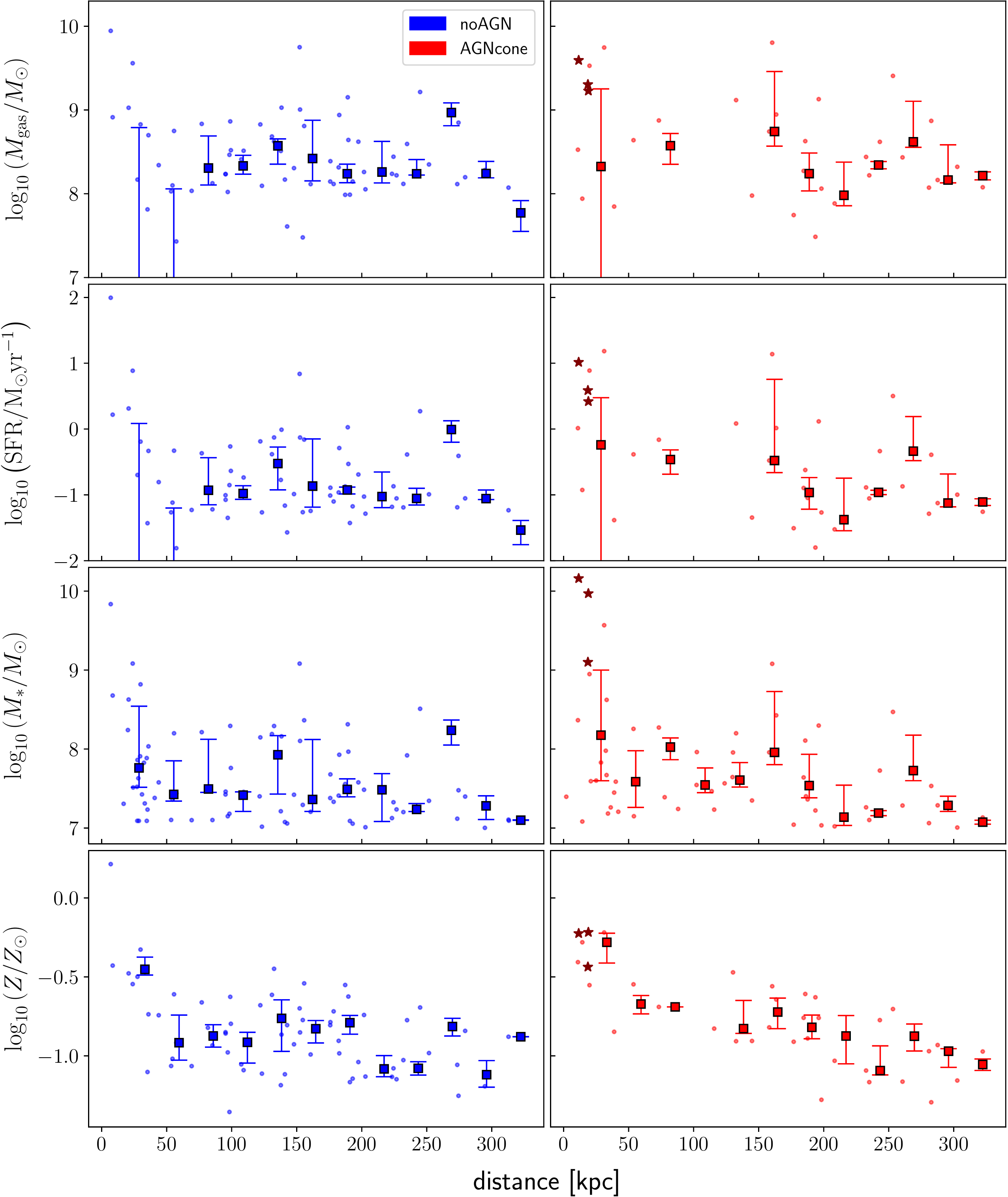}
    \caption{Satellites intrinsic properties as a function of the distance from the accretion-weighted centre ${\bf c}_{aw}$ in the runs \noAGN{} ({\it left column}) and \AGNcone{} ({\it right column}) at $z=6$. The main, central galaxy is not shown. From top to bottom: gas mass, stellar mass, star formation rate, and metallicity. Brown stars refer to the most active AGN, here defined as those galaxies hosting at least a MBH with a total accretion rate ${\dot{M}_{\rm acc}>1}$~$M_{\sun} yr^{-1}$. The median of the distributions are superimposed in each panel as filled squares. Uncertainties are quantified as the 30th and 70th percentiles of each distribution.
    The values of the median points and the lower bounds of the errorbars not shown in the plots are equal to zero.}
    \label{fig:intrinsic_vs_dist}
\end{figure*}

To start investigating the effect of AGN feedback on the quasar environment, we compute the redshift evolution of the number of companions, as shown in Figure~\ref{fig:gal_count}.
The three families of lines refer to those objects enclosed within different spheres, centered on the ``accretion-weighted center", ${\bf c}_{aw}$ (see Appendix~\ref{sec:AccretionCentre}), and with increasing radii: $1$ central dominant (cD) galaxy virial radius (about $66$~kpc at $z=6$), 3 $r_{\rm vir}$, and 5 $r_{\rm vir}$.

The number of satellites in the two simulations (red and blue lines) is very similar at any redshifts within 1 virial radius, starting from a few sources at $z=10$ and reaching about 10 objects at $z=6$.
At larger distances from the cD, an increasing difference between the runs starts to arise: for $z\lesssim7$, the number of satellites in the \AGNcone{} case is smaller than in the \noAGN{} one.
After $z\sim7$, i.e. when the outflow starts affecting the properties of the host galaxy\footnote{See Figure~8 in \citealt{Barai_et_al_2018}: $\dot{M}_{\rm out}$ increases substantially after $z\sim 8$ near the cD virial radius, and by $z\sim7$, the outflows reach the outskirt of the cluster.}, the number of satellites in the outer regions of \AGNcone{} becomes smaller than the number in \noAGN{}, whereas a substantial agreement is maintained at smaller radii.
For $r>r_{\rm vir}$, \AGNcone{} satellite number reaches a peak at $z\sim6.7$ and is constantly reduced at lower-$z$.
Differently, in \noAGN{} the satellite number is always a decreasing function of redshift.
The highest discrepancy of 24 satellites between the two simulations is reached at $z=6$, within 5 $r_{\rm vir}$.
\subsection{Redshift evolution: satellite properties}
\label{subsec:redshift_evolution}

Figure~\ref{fig:redshift_evolution} shows the redshift evolution ($6 \lsim z \lsim 10$; for a sub-sample of snapshots) of several satellite\footnote{In each snapshot, we exclude from the computation the most massive galaxy in the two runs. Even if the proper way to identify a quasar companion would require to identify those galaxies which orbit around an accreting MBH host (see the method adopted in \ref{subsec:EAGN}) in \AGNcone{} and to look for their counterpart in \noAGN{}, here our approximation is equivalent for the analysis.} properties: gas mass ($M_{\rm gas}$), star formation rate (SFR), stellar mass ($M_{*}$), and gas metallicity ($Z$) in solar units.
The median of the gas content of our satellite population slowly decreases with time, from $\sim6\times 10^{8}~\msun$ to $\sim10^{8} \msun$ in about $0.5$~Gyr. As a consequence, the capability to form stars of the satellites decreases: the median SFR, in fact, varies from $\sim$1 to $0.1~\msun$~yr$^{-1}$.
The median stellar mass fluctuates around $M_{*}\sim3 \times10^{7}~\msun$, with a shallow increase from $z=10$ to $z=6$. The ISM is consequently gradually enriched in metals as stars are formed and explode as SNe, varying between $\sim0.05$ and $0.1$~$Z_{\odot}$.

The differences between the runs are minimal and well within their error bars, thus suggesting that AGN feedback only plays a minor role in shaping the overall evolution of satellite properties.
We notice however that the values of $M_{\rm gas}$ and SFR are higher in \AGNcone{} till $z=7$, whereas $M_{*}$ and $Z$ are almost always larger in \AGNcone{} with respect to \noAGN{}.  

\subsubsection{Stellar mass-metallicity relation}
\label{subsubsec:mstar-met}

In Figure~\ref{fig:Mstar_Z}, we compare our results with observational data, concerning the $M_*-Z$ relation in isolated galaxies. 
In particular, we show the gas-phase metallicity from the oxygen vs hydrogen abundance ratio for the redshifts $z=6,7,8$, along with a linear fit of all these satellite populations.
As suggested by observations, also in our simulations, systems with increasing stellar mass are progressively more polluted in metals. The normalization of the sub-linear $M_*-Z$ relation increases with decreasing redshift indicating that the overall metallicity floor of the galaxy group increases through cosmic times. 

Similarly to the other intrinsic properties, the general trends of the run \noAGN{} are almost indistinguishable from the \AGNcone{} results.
If we linearly interpolate the $M_*-Z$ distributions at $z=6$, we notice that the slope of the best-fit relation is only slightly steeper in \noAGN{} than in \AGNcone{}, being $0.33\pm0.03$ versus $0.25\pm0.05$, thus consistent with the errors.
This indicates that the AGN feedback does not change significantly the process of gas pollution in the galaxy group as a whole.
However, we also note that the difference between the zero-points is marginally larger, with $5.6\pm0.2$ versus $6.2\pm0.3$, which is compatible with a higher SF activity in the AGNcone past evolution.

The comparison between observations and our results is however not straightforward.
Due to sensitivity limitations, observations are still unable to probe the low-mass end ($M_*\lsim 10^9~\msun$) of the relation, that is instead statistically covered by our simulated data.
Future deeper observational surveys are therefore necessary to reduce this bias.
In general, we note that the level of metal enrichment reached by the galaxy groups of both runs seems to agree with the extrapolation from the data of $z=6$ galaxies \citep{Harikane_et_al_2020} (green crosses).

\subsection{Spatial distribution of companions}

In this Section, we analyse how satellite properties are spatially distributed within the galaxy group, to check any possible correlation with the AGN activity, occurring in the \AGNcone{} simulation.

The kinetic feedback, as modelled in our simulations, might remove gas from galaxies close to ${\bf c}_{aw}$ and transfer it to more peripheral systems. The way it finally affects the gas distribution of the galaxy group is anyway complex. In the dark matter distribution, high density peaks tend to be clustered \citep{Bardeen_et_al_1986}, implying that most massive halos are closer to the center of mass of the galaxy proto-cluster (``mass segregation"). Since the center of mass almost coincides with ${\bf c}_{aw}$ at $z=6$, one should expect the effect of quasar feedback to be higher in the closest, more massive satellites. However, as a consequence of their deeper potential wells, massive systems might retain their gas content more easily with respect to less massive ones, thus being more resilient to the possible passage of outflows launched from the galaxy itself or coming from close companions.

To study the effect of quasar feedback on the spatial distribution of satellites, in Figure~\ref{fig:intrinsic_vs_dist} we report the satellite intrinsic properties ($M_{gas}$, SFR, $M_*$, $Z$) as a function of their distance from ${\bf c}_{aw}$.
Both runs show a highly dense environment, where the most gas- and metal-rich, star-forming, galaxies are preferentially located at small distances from the center of the galaxy group, independent of the presence of quasar feedback. 
This suggests that the satellite distribution within the galaxy proto-cluster is dominated by mass segregation and quasar feedback plays, at most, a second-order effect.
In general, we find the distribution of satellites to be quite flat at distances larger than $100$~kpc.

Still, in \AGNcone{} there is a larger number of star-forming ($SFR\gsim1~\msun~yr^{-1}$) massive ($M_*\gsim10^9~\msun$) galaxies with respect to the \noAGN{} run. This result, along with the one reported in \S~\ref{subsec:redshift_evolution} (larger values of $M_{\rm gas}$, $M_{*}$, SFR, and $Z$ in \AGNcone{} with respect to \noAGN{}), suggests to refine the comparison between the runs by focussing on individual satellites. In fact, the effect of quasar feedback on the environment could be washed-out by averaging the satellites properties over the entire population. Different galaxies may indeed perceive feedback effects at different times, depending on their relative position with respect to the most accreting MBH and on the variability of the MBH itself. 


\section{Individual companions}
\label{sec:effect_individual}

\begin{figure*}
    \includegraphics[width=\textwidth]{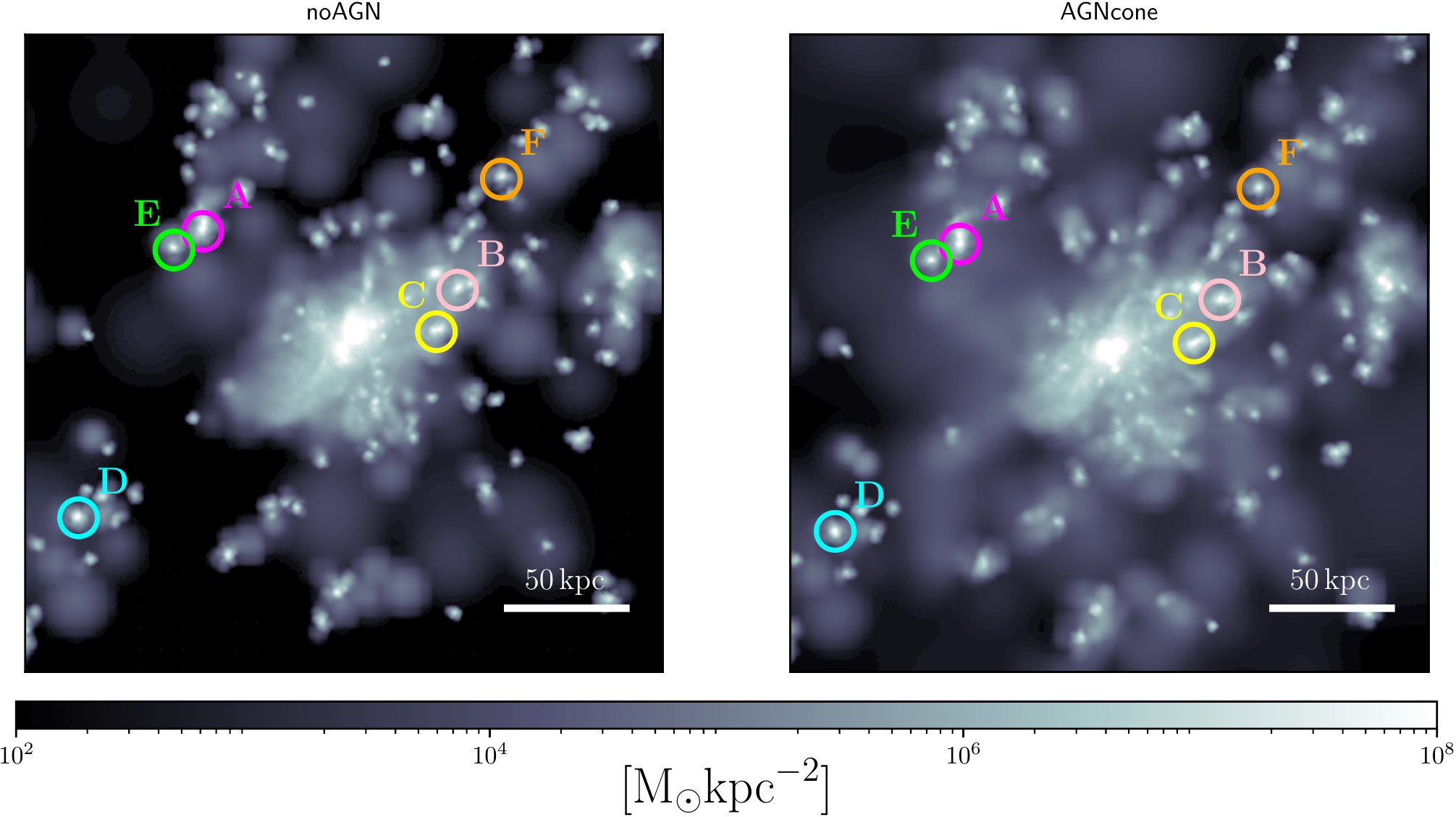}
    \caption{$z=6.3$ stellar surface density maps of the central $270$~kpc of \noAGN{} ({\it left panel}) and \AGNcone{} ({\it right panel}).
    Matched galaxies are highlighted in the panels with a circle of the same colour.}
    \label{fig:stars_match}
\end{figure*}

\begin{figure*}
    \includegraphics[width=\textwidth]{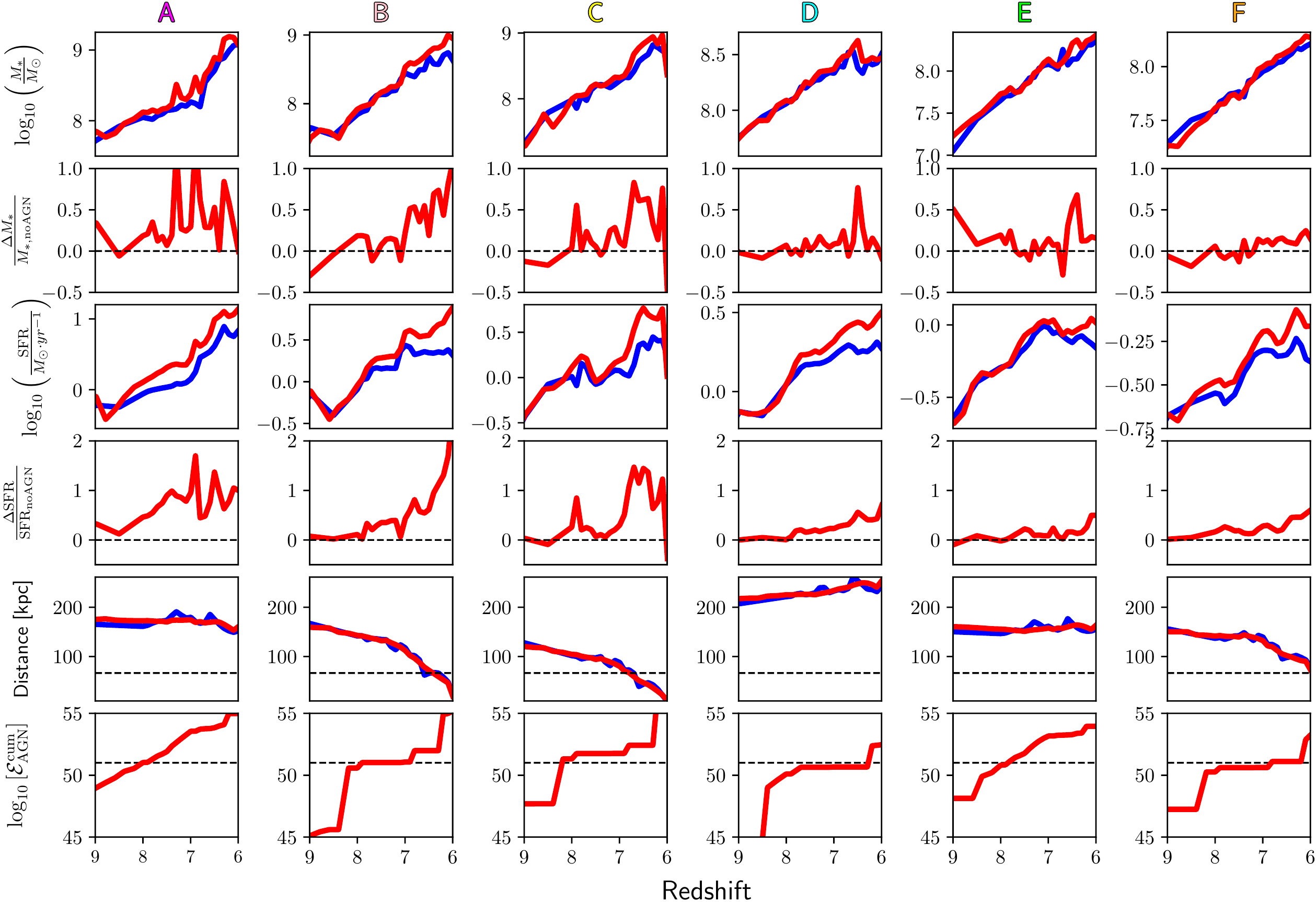}
    \caption{Redshift evolution of the six galaxies in the sample, as described in Section~\ref{sec:effect_individual}.
    From top to bottom: stellar mass, stellar mass relative difference (i.e. $[M_{*, \rm AGNcone}-M_{*, \rm noAGN}]/M_{*, \rm noAGN}$), SFR, SFR relative difference (i.e. $[\rm{SFR}_{\rm AGNcone}-\rm{SFR}_{\rm noAGN}]/\rm{SFR}_{\rm noAGN}$), distance from ${\bf c}_{aw}$, and $\mathcal{E}_{\rm AGN}^{\rm cum}$.
    The same colour coding of Figure~\ref{fig:redshift_evolution} is applied for the two runs.
    Horizontal lines mark the zero-level for the relative differences (second and forth rows), the time-averaged position of the cD virial radius (fifth row), and the threshold $10^{51}$~erg for the cumulative integrated flux (solid lines in sixth row; see \S~\ref{subsec:EAGN}).}
    \label{fig:Gal_matching_SFR_Mstar}
\end{figure*}

The goal of this Section is to compare the SFR and $M_*$ redshift evolution of a satellite in the \AGNcone{} run with the corresponding satellite in the \noAGN{} run. Finding the same satellite, available in both runs in the same redshift interval is not trivial, since different feedback prescriptions result into different merger rates (see \S~\ref{subsec:redshift_evolution_number}): in \AGNcone{}, galaxies merge faster and more easily than in \noAGN{}, possibly because of a more diffuse and massive stellar component around the galaxies (see Figure~\ref{fig:stars_match}).
We thus describe in details the procedure we follow to setup our sample of galaxy couples.

We select a sample of galaxies that are characterised by ($i$) the same position (within $5$~comoving kpc) and ($ii$) the same virial mass (within $10$\%) in both runs. Among the possible candidates from this first selection ($iii$) we select galaxies that, at $z=6$, have different distances from ${\bf c}_{aw}$ and different masses.
This condition allows us to probe different regions of the quasar environment in terms of mass segregation. Furthermore, we choose galaxies whose ($iv$) merger tree starts at least\footnote{This  requirement takes out from the selection those objects which suffer from violent gravitational stripping processes or merge with a galaxy of equal/higher mass, long before its counterparts in the other run.} at $z=9$ and reaches $z=6$. Finally, ($v$) we exclude from our sample those galaxies that have hosted a powerful AGN for a significant amount of their evolutionary history.

\noindent Although only a small fraction of satellites hosts a powerful AGN, this last conditions is essential to isolate the effect of external from internal AGN feedback on satellites. AGN-driven outflows may, in fact, subtract part of the cold gas component from the interstellar medium of the host galaxies, affecting their SFRs and stellar masses.
To select galaxies hosting powerful AGN, we first walk backward the merger tree of each galaxy (selected at redshift $z$) and then we sum, at each redshift, up to the formation redshift\footnote{We define the formation redshift $z_{\rm form}$ of a galaxy, as the first redshift (the highest) where the first ancestor of the galaxy in the merger tree is identified.} $z_{\rm form}$, the accretion rate ${\dot{M}_{acc}}$ of all the MBHs located within $\beta r_{\rm vir}$. 
After this, we compute the mean $\langle\dot{M}\rangle_{acc}$ of the cumulative accretion rate of the selected galaxy over the whole redshift range and exclude from the analysis those galaxies for which ${\langle\dot{M}\rangle_{\rm acc}>0.1}$~$M_{\sun}$~yr$^{-1}$. 
Via this method, we always exclude at least the cD galaxy, hosting some of the most active and massive MBHs during the whole evolution.\footnote{We note that, in principle, this criterion might miss MBHs that are highly accreting at low mass, possibly neglecting relevant effects in the case the MBH is hosted in a small mass galaxy.}

The final selection, composed by 6 galaxies with virial masses\footnote{With an average contamination from low-resolution DM particles of a few percent for 1 out of the 6 galaxies and null for the remaining 5 objects.} $\sim 10^{10}-10^{11}~\msun$, is highlighted in Figure~\ref{fig:stars_match}; in Figure~\ref{fig:Gal_matching_SFR_Mstar}, we show the redshift evolution of stellar mass, SFR and distance from ${\bf c}_{aw}$ for each galaxy of the sample.
At higher redshift the values of $M_{*}$ and SFR (first and third rows) are almost identical in the two simulations, for all the galaxies.
After $z\sim8$, the trends start to diverge, always resulting in higher stellar mass and SFR in \AGNcone{} satellites (the three galaxies on the left - i.e A, B, and C - have larger differences with respect to the three rightmost galaxies D, E, and F, as it will be discussed in \S~\ref{subsec:EAGN}).
The sudden fluctuations observable in the evolution of some satellites (e.g object A at $z\sim7$) are due to minor mergers or close fly-bys which temporary increase the mass within $\beta r_{\rm vir}$.
The abrupt decrease of the red line of C is due to the upcoming coalescence of the object with the cD, at the very last snapshot.
Hence, C system is majorly stripped of both its gaseous and stellar components. 
\noindent The increasing differences are more easily noticeable in the relative difference panels (second and forth rows), where positive values represent higher quantities in \AGNcone{} run.
As it is clear from the fifth row, satellite distances evolve quite differently among the selected objects, further increasing the generality of the sample: while B, C, and F approach the centre, galaxy D orbits increasingly far form the cD, and galaxies A and E do not significantly change their relative position during the entire simulation time.

\begin{figure*}
    \includegraphics[width=0.8\textwidth]{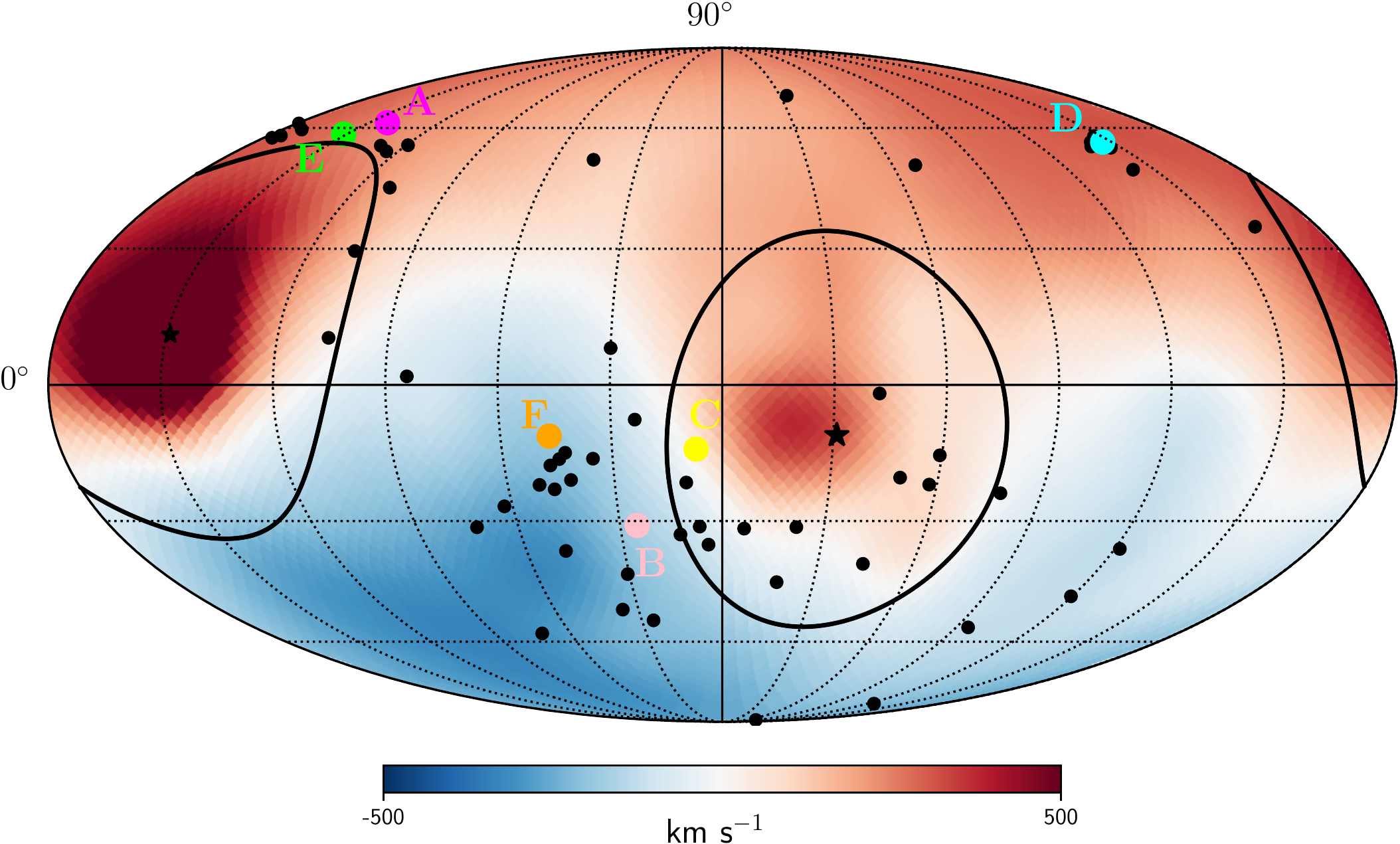}
    \caption{Mollweide view of the gas radial velocity within $10$~kpc of the most accreting MBH in \AGNcone{}, at $z=6.3$. The map shows the effect of the bi-conical feedback, resulting in two visible gaseous outflows. In the \AGNcone{} run, the orientation of the outflow is randomly assigned to each MBH at the seeding time, and kept fixed for the whole MBH lifespan. The gas velocity distribution, however, also depends on the activity of the other MBHs present in the simulation. The direction of the emission axis for the most accreting MBHs is recovered by fitting the radial velocity map of the gas. The black stars mark the position of the opposite outflow axes, as they are derived from the velocity distribution, whereas the black lines represent the boundaries of the outflows, ideally confined within $45^{\circ}$ from the axis. Black dots show the angle position of the satellites at $z=6.3$. Larger coloured dots mark the galaxies selected for the matching study of Section~\ref{sec:effect_individual}}. The colour coding is the same used in Figure~\ref{fig:stars_match}.
    \label{fig:mollview}
\end{figure*}

To visualise the relative position of the selected satellites with respect to the geometry of the bi-conical outflows, we show in Figure~\ref{fig:mollview} a Mollweide map, where the location of each galaxy at $z=6.3$ is superimposed to the radial velocity map of the gas within a sphere of radius $10$~kpc and centred on the most accreting MBH in that snapshot.

\subsection{AGN feedback on individual companions}
\label{subsec:EAGN}

Given the non-isotropic emission of the out-flowing gas in \AGNcone{}, the evaluation of the feedback effect cannot rely only on the distance from the source and the target galaxy.
In addition, galaxies randomly experience the influence of various AGN during various and discontinuous accretion periods.
It is then necessary to define a function of both distance and emitted power that can also take into account the relative position of a target galaxy with respect to the emission axes.

\begin{figure}
    \includegraphics[width=0.48\textwidth]{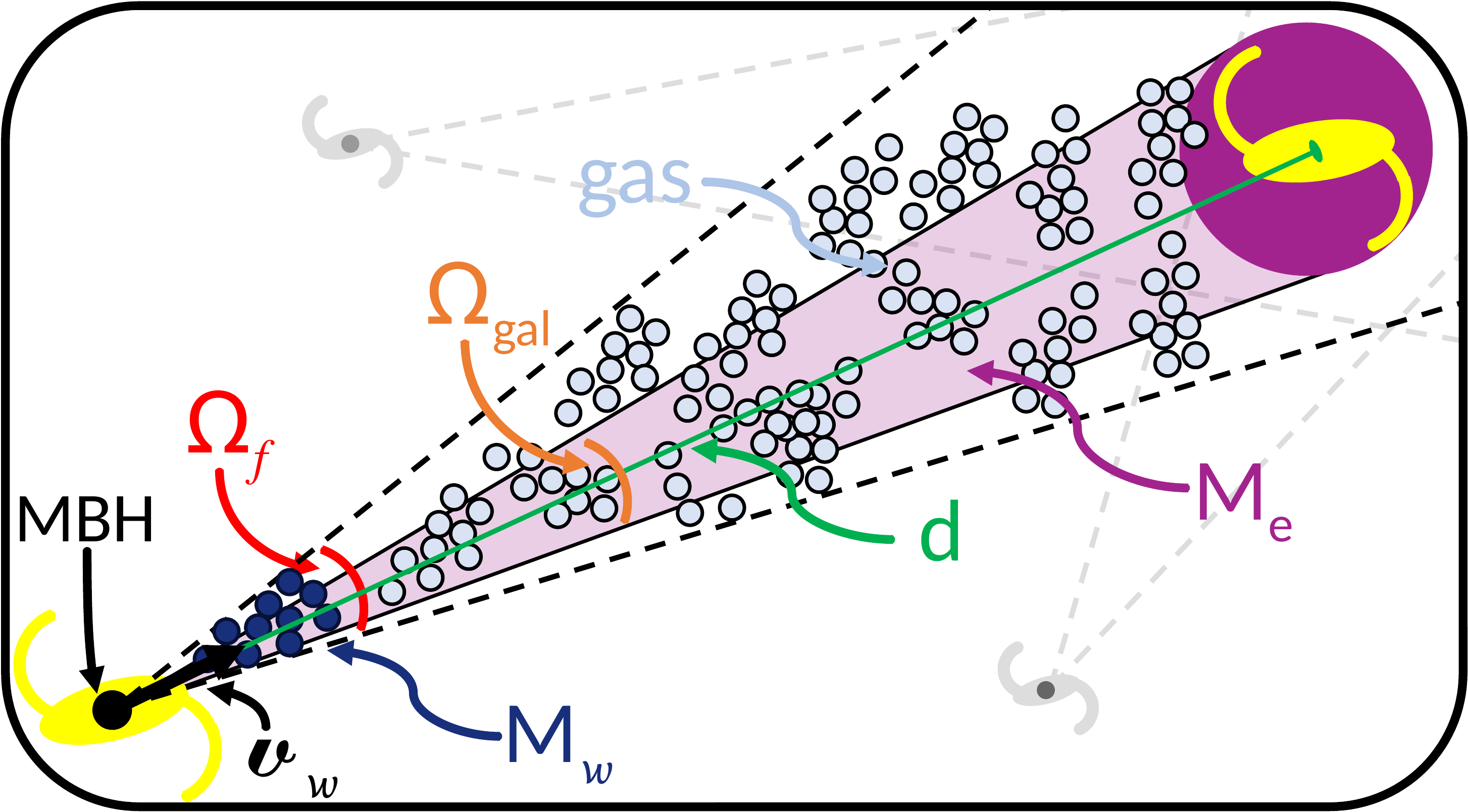}
    \centering
    \caption{Schematic representation of the model applied to \AGNcone{}. From an accreting MBH, a gaseous outflow with mass $M_{\rm w}$ is ejected with velocity $v_{\rm w}$ in the cone with aperture $\Omega_{f}$ and intercepts a satellite at the distance $d$, subtending an angle $\Omega_{\rm gal}$. $M_{e}$ is the gas envelope mass encountered by the outflow when it is ejected.}
    \label{fig:toymodel}
\end{figure}

Let us consider a galaxy (top right in Figure~\ref{fig:toymodel}) at a distance $d$ from an accreting MBH (bottom left), subtending an angle $\Omega_{\rm gal}$.
The MBH launches an outflow (dashed lines) towards the galaxy in an opening angle
$\Omega_{\rm f}$ (identified by the dotted lines). 
We quantify the energy received by the target galaxy at a given redshift $z$, from any $i$-th MBH which has been accreting with $\dot{M}_{{\rm acc}, i}$ in the time interval $\Delta t_{\rm snap}$, and whose outflow cone encompasses the galaxy centre.\footnote{In the following derivation, we neglect the subscript $i$ on the MBH properties to lighten the formalism.}

The total gas mass involved in the feedback emission process around the MBH is $M_{\rm tot}=M_{w}+M_{\rm e}$, where $M_{\rm w}$ is the mass of the ejected wind and $M_{e}$ is the mass of the gas in the environment surrounding the MBH and entrained within the outflow.
Depending on the geometry of the feedback prescription, $M_{e}$ can be written as
\begin{equation}
    M_{e} = \frac{\Omega_{\rm f}}{4\pi}\frac{4\pi}{3}\rho d^3,
    \label{eq:menv}
\end{equation}
where $\rho$ is the average density around the MBH.
If $v_{\rm f}$ is the velocity of the gas when it hits the target galaxy, then momentum conservation\footnote{We assume that the energy of the outflow can be radiated away during the gas migration, because of gas heating and shock fronts.} allows us to write
\begin{equation}
    M_{\rm w}v_{\rm w}+M_{\rm e}v_{\rm e} = v_{\rm f}M_{\rm tot}
    \label{eq:momcons}
\end{equation}
where we assume $v_{\rm e}=0$ when the outflow is launched.

Neglecting internal energy\footnote{The energy of the outflow is almost completely kinetic, also according to the feedback receipt.}, the energy deposited by the i-th MBH on the target galaxy ($\mathcal{E}_{\rm AGN, i}$) is related to the final kinetic energy $E_{\rm f}$ through the relation:
\begin{equation}
    \mathcal{E}_{\rm AGN, i} \equiv E_{\rm f} \frac{\Omega_{\rm gal}}{\Omega_{\rm f}} = \frac{1}{2}M_{\rm tot}v_{\rm f}^{2} \frac{\Omega_{\rm gal}}{\Omega_{\rm f}},
    \label{eq:Edep1}
\end{equation}
where the factor $\Omega_{\rm gal}/\Omega_{\rm f}$ accounts for the fact that only a fraction of the total mass ejected by the MBH actually intercepts the target galaxy.

Isolating $v_{\rm f}$ from Equation~\ref{eq:momcons} and considering that the envelope mass in Equation~\ref{eq:menv} fully dominates over the wind mass (i.e. $M_{\rm tot} \simeq M_{\rm e}$), Equation~\ref{eq:Edep1} becomes
\begin{equation}
    \mathcal{E}_{\rm AGN, i} = \frac{3}{2}v_{\rm w}^2 \frac{M_{\rm w}^2}{\rho d^3} \frac{\Omega_{\rm gal}}{\Omega_{\rm f}^2}.
    \label{eq:Edep2}
\end{equation}
In Equation~\ref{eq:Edep2}, the solid angle $\Omega_{\rm gal}$ -- subtended by the target galaxy with respect to the MBH position -- can be approximated as
\begin{equation}
    \Omega_{\rm gal} = \frac{A_{\rm gal}}{d^2},
    \label{eq:SA}
\end{equation}
where $A_{\rm gal}$ is the projected area of the galaxy, as it is seen from the AGN.\footnote{We note that, the correct formula for a curved surface subtending the solid angle $\Omega_{\rm gal}$ is $\Omega_{\rm gal}=2\pi(1-\frac{d}{\sqrt{r_{\rm gal}^2+d^2}})$. Our approximation fails only when $d \sim r_{\rm gal}$, which occurs a negligible amount of times, leading our method to overestimate the solid angle, in those few cases, at most by a factor 1.7.}
In our analysis we use $A_{\rm gal} = \pi \beta^2 r_{\rm vir}^2$, with $\beta = 0.3$, in agreement with our choice to compute the properties of the halos (see Section~\ref{sec:sample}).
At the same time, $\Omega_{\rm f} = 2\cdot2\pi(1-\cos \alpha)$, with $\alpha = \pi/4$ in \AGNcone{}, for each of the two cones where the energy is distributed.

In order to estimate $M_{\rm w}$, we multiply the outflow rate $\dot{M}_{\rm w}$ into the snapshot time interval $\Delta t_{\rm snap}$:
\begin{equation}
    M_{\rm w} = \dot{M}_{\rm w} \Delta t_{\rm snap},
    \label{eq:Mdot_Deltat}
\end{equation}
where $\dot{M}_{\rm w}$ is obtained from the accretion rate ${\dot{M}_{acc}}$ via equations~\ref{eq:Lrad} and \ref{eq:Efeed}.
Accordingly, the energy-conservation equation can be written as
\begin{equation}
    \dot{E}_{f} = \frac{1}{2}\dot{M}_{\rm w}v_{\rm w}^2 = \epsilon_{\rm f}\epsilon_{\rm r}\dot{M}_{\rm acc}c^2.
    \label{eq:Erad1}
\end{equation}

Finally, we can write down the form of the deposited energy, by substituting $M_{\rm w}$ in Equation~\ref{eq:Edep2}:
\begin{equation}
    \mathcal{E}_{\rm AGN, i} = 6 \pi \epsilon_{\rm r}^2 \epsilon_{\rm f}^2 c^4 \beta^2 \frac{(\Delta t_{\rm snap})^2}{v_{\rm w}^2} \frac{\dot{M}_{\rm acc}^2 r_{\rm vir}^2}{\Omega_{\rm f}^2 \rho d^5}.
    \label{eq:Edep3}
\end{equation}

Operatively, we evaluate the gas density $\rho$ by summing up all the gas particles in the cone subtended by the solid angle $\Omega_{\rm gal}$, with respect to the $i$-th MBH and the accretion rate $\dot{M}_{\rm acc}$ at the time $t-\Delta t_{\rm snap}$, in agreement with what is done in Appendix~\ref{sec:AccretionCentre}.

The derived Equation~\ref{eq:Edep3} has a quite intuitive dependence on the MBH accretion rate, $\dot{M}_{\rm acc}^2$ (the stronger the AGN, the higher the effect) and on the size of the galaxy, proportional to $r_{\rm vir}^2$ (the larger is the galaxy projected area with respect to the AGN, the higher is the energy harvested by the system). Also the dependencies on the inverse of the distance galaxy-MBH, $d^5$ and on the density of the circumgalactic medium (CGM) are expected. As a matter of fact, the density of the environment and the length of the path that the outflow has to travel both contribute to lower the final energy. In a purely momentum-conservation scenario the kinetic energy of the ejected gas is continuously distributed on the surrounding medium and, in very dense environment, can be completely lost in the CGM before reaching the target system. A final consideration concerns the solid angle $\Omega_{\rm f}$, which favours more collimated outflows. This factor, on one hand is able to reduce the volume where the feedback energy is diluted but, on the other, it deeply affect the probability that a galaxy is affected at all by the outflow.

In order to quantify the total effect of the MBH population on each galaxy, we define the total $\mathcal{E}_{\rm AGN}$ as
\begin{equation}
    \mathcal{E}_{\rm AGN} = \sum_{i=1}^8 \mathcal{E}_{{\rm AGN}, i},
\end{equation}
where the summation is carried over the eight most accreting MBHs as discussed in Appendix~\ref{sec:AccretionCentre}.
In our model, for a given target galaxy, we consider the contribution of the $i$-th accreting MBH only if the galaxy centre of mass is enclosed within the boundaries of the outflow launched by the AGN (see the black solid lines in Figure~\ref{fig:mollview}).
This equation provides a rough estimate of the energy deposited in a time interval $\Delta t_{\rm snap}$ by the most-accreting AGN on a target galaxy, at redshift $z$.

At any generic redshift $z$, the cumulative energy deposited on a target galaxy ($\mathcal{E}_{\rm AGN}^{\rm cum}$) is given by the integral of $\mathcal{E}_{\rm AGN}$ between the galaxy formation redshift $z_{\rm form}$, and $z$:
\begin{equation}
    \mathcal{E}_{\rm AGN}^{\rm cum} \equiv \kern-1em \sum_{\quad z=z_{\rm form}}^{z} \kern-1em \mathcal{E}_{\rm AGN}.
    \label{eq:Edep4}
\end{equation}

The results of this model are shown in the bottom line of Figure~\ref{fig:Gal_matching_SFR_Mstar}, where the horizontal line, at $\mathcal{E}_{\rm AGN}^{\rm cum}=10^{51}$~erg, refers to the energy injected in the medium by a single SN event and has the only purpose of helping the reader to distinguish the objects in relation to their received energy.
According to the values of $\mathcal{E}_{\rm AGN}^{\rm cum}$, we separate the galaxies more affected by the AGN feedback (A, B, C, on the left) from the less affected ones (D, E, F, on the right).
The SFR of galaxies from the first group is more enhanced (up to a factor 3) with respect to the second group (less than a factor 2).

We note that in A, B, and C, $\mathcal{E}_{\rm AGN}^{\rm cum} \gtrsim 10^{51}$ from $z\sim 8$ up to the end of the simulation, while in D, E, and F $\mathcal{E}_{\rm AGN}^{\rm cum} \gtrsim 10^{51}$ only for $7.5<z<6$.
In other words, it seems that galaxies receiving the energy by the AGN feedback for a longer time are more strongly affected, in terms of SFR and $M_*$. 
We discuss in the next section a possible interpretation of these findings.


\section{Interpretation}
\label{sec:discussion}
%
In this section, we first summarise the main findings obtained from the comparison between the runs \noAGN{} and \AGNcone{}, and we propose our interpretation for our results.

In both runs, we have identified a sample of galaxies in the redshift range $6<z<10$ and characterized their properties (distance from the center of the galaxy groups, SFR, $M_*$, $M_{\rm gas}$, $Z$).
From the comparison between the samples extracted from the two simulations we can conclude that:
\begin{enumerate}[label=(\alph*)]
    \item \label{itm:sat_num} In the \AGNcone{} run satellites are less numerous, especially in the outer regions, and they are hosted by more massive DM halos (see Section~\ref{sec:sample});
    \item \label{itm:sat_sfr} Although the differences between the median properties of the two samples at all redshifts are not noteworthy, individual \AGNcone{} companions are more star-forming and massive; the SFR enhancement in those satellites which are influenced the most by the surrounding active AGN, reaches a factor up to 4 (see Sections~\ref{sec:sample} and \ref{sec:effect_individual}).
\end{enumerate}

We suggest that point \ref{itm:sat_num} is due to the different merger rates in the two simulations. Figure~\ref{fig:stars_match} shows the presence of a diffuse stellar component in the proto-cluster environment of \AGNcone{}, absent in \noAGN.
This is due to the combined activity of the dimly accreting MBHs, hosted in the satellites orbiting in the outskirt of the group, and to the powerful quasars located at the centre.
The disperse gas and stars can boost the effect of dynamical friction when two galaxies approach, thus lowering the dynamical timescale for a merging to occur. DM-halos follow a similar trend, suggesting that differences in the baryonic component are transferred to DM structures that, in the \AGNcone{} run, become more massive, and therefore less in number.
Interestingly, the effect is only relevant far from the cD and almost absent near the centre, where the dynamics is completely dominated by the cD environment. Strong feedback could intrinsically favour coalescence, decrease the number of systems, and speed up their bottom-up growth. These phenomena play a fundamental role in the number count of satellites around $z=6$ quasars, as discussed in the next section.

Concerning point \ref{itm:sat_sfr}, we note that AGN feedback do not significantly affect the evolution of the satellite population as a whole.
The median of the satellite properties show only minor differences, which are though compatible with a scenario where, in \noAGN{}, a higher $M_{\rm gas}$ results in a higher SFR, till the exhaustion of the gas reservoirs. We note that, the cumulative effect on the SFR and $M_{*}$ is both due to the higher merger rate (more likely to occur far from the cD) and to the direct effect of the outflow on the satellite (stronger at smaller distances from the cD).
We suggest that the enhanced SF in those satellites directly invested by the AGN outflows can be explained by ($i$) an increase of the fuel available for the SF process, ($ii$) a boost in its efficiency due to the induced shocks, ($iii$) the formation of stars within the outflow itself \citep[][]{Maiolino_et_al_2017,Gallagher_et_al_2019}.

Several evidences of AGN positive feedback onto galaxy satellites have been reported in the literature. \citet{Gilli_et_al_2019} analyse the deep multi-band field of a type II radio galaxy at $z=1.7$ and discover an over-density of galaxies. These authors suggest that the star formation in satellites is promoted by the compression of their cold interstellar medium around the AGN-inflated bubbles. The possibility of a jet-induced star formation is also consistent with the SF measurement of a galaxy near the local radio-galaxy NGC--541 \citep{Croft_et_al_2006}, and supported by dedicated numerical simulations \citep{Fragile_et_al_2017}. 

Recently, \citet{Martin-Navarro_et_al_2021} measured $M_*$ and SFR in a large sample of satellites in SDSS galaxies finding that their SFR is modulated by their relative position with respect to the central galaxies, being higher in those objects presumably affected by the outflows from the central galaxies.
These authors concluded that AGN-driven outflows can influence the environment well beyond the host galaxy ($1-2~R_{\rm vir}$), preserving (or even enhancing) the SF of those satellites along the direction of the injected energy.

There are some unavoidable limits associated with our model to quantify $\mathcal{E}_{\rm AGN}$.
On the one hand, being based only on momentum conservation, it would underestimate the final energy deposited into the target galaxy by a fully energy-conserving outflow. On the other hand, the outflow launched by AGN may interact with intervening ISM and CGM material, potentially losing a significant amount of energy, or even being stopped in the most extreme cases. This would instead lead us to overestimate the final impact on the satellite.

In addition, the limited spatial and temporal resolution of the simulation suite, the random explosion of SNe, and the possible accretion episodes of MBHs within the satellites themselves\footnote{We remark though that we limit the impact of AGN hosted in satellites by excluding those objects with ${\langle\dot{M}\rangle_{\rm acc}>0.1}$~$M_{\sun}$~yr$^{-1}$.}, may dilute the effect of feedback from external AGN.
All these caveats prevent us from disentangling entirely the effects of the several physical processes affecting the CGM and enhancing the SF in satellite galaxies.


\section{Comparison with observations}
\label{sec:observations}

\begin{figure*}
    \centering
    \includegraphics[width=\textwidth]{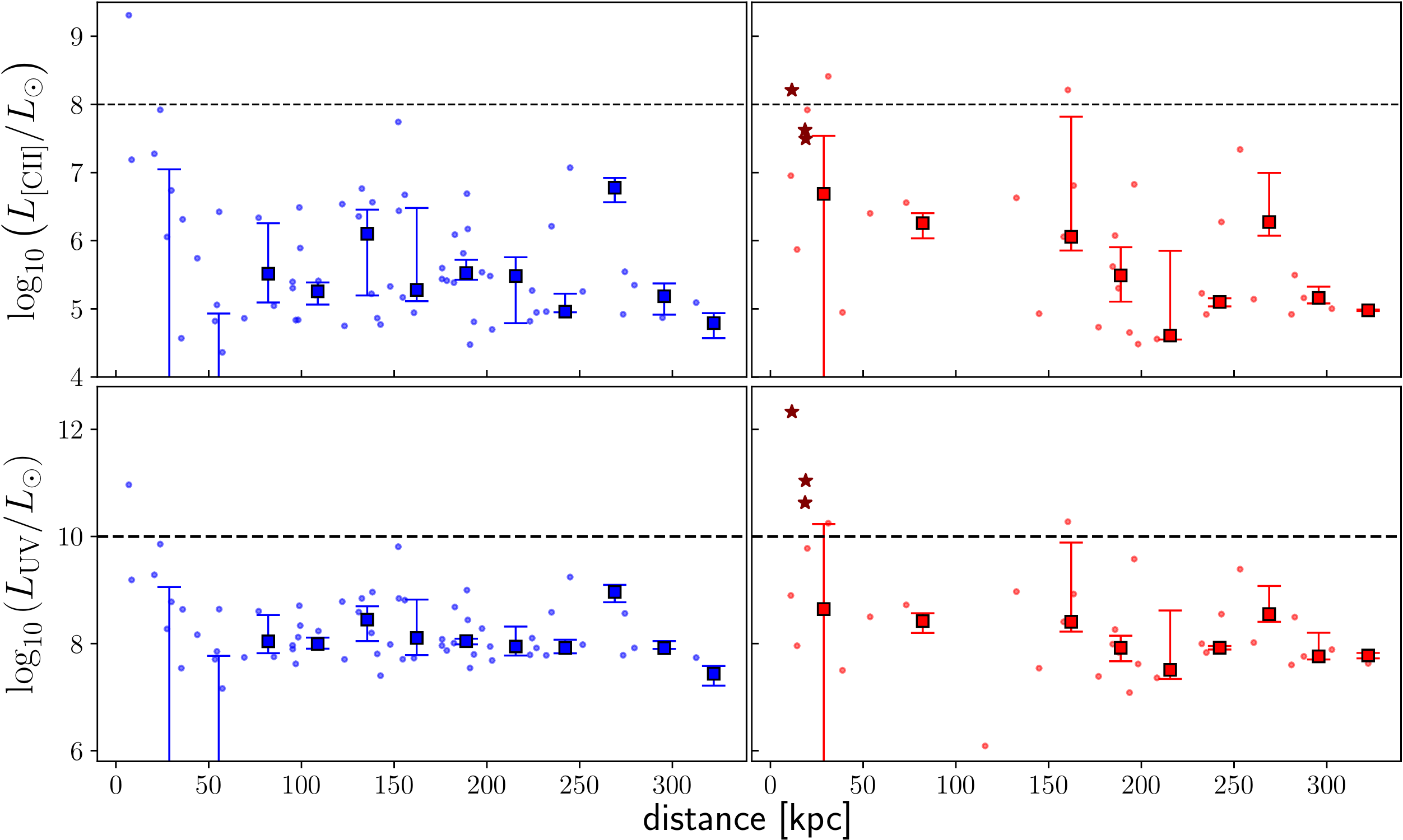}
    \caption{[CII] emission (top row) and UV emission (bottom row) as a function of distance from ${\bf c}_{aw}$, for the same $z=6$ sample of Figure~\ref{fig:intrinsic_vs_dist}: \noAGN{} on the {\it left} and \AGNcone{} on the {\it right}.
    Horizontal dashed lines provide an estimate of the current instrumental sensitivities, derived from some of the deepest observations in the various bands.
    As in Figure~\ref{fig:intrinsic_vs_dist}, where the brown stars mark the hosts of the most active AGN in \AGNcone{}, the filled squares show the medians of the distributions and the error bars refer to the 30th and 70th percentiles.
    The values of the median points and the lower bounds of the errorbars not shown in the plots are equal to zero.}
    \label{fig:observative_vs_dist}
\end{figure*}

In order to evaluate how AGN feedback affects our capability to detect an over-density through the number counts of galaxies, we compute both the [CII] and UV luminosity of all the simulated satellites as detailed in Appendix~\ref{sec:observational_properties}.
Figure~\ref{fig:observative_vs_dist} shows the satellite luminosity as a function of the distance from the center of the system at $z=6$.
Horizontal dashed lines mark the observed limit luminosities $L_{\rm lim}$ of current observational campaigns at high redshift: $L_{\rm lim, [CII]} = 10^{8} L_{\odot}$ \citep[][hereafter \citetalias{Venemans_et_al_2020}]{Venemans_et_al_2020} and $L_{\rm lim, UV} = 10^{10} L_{\odot}$ \citep[][]{Marshall_et_al_2020}. 
In agreement with our previous results, \AGNcone{} produces more luminous, and thus more easily detectable, galaxies than the control simulation.
In details, above $L_{\rm lim, [CII]}$ ($L_{\rm lim, UV}$), there are 1 and 3 (1 and 5) satellites in \noAGN{} and \AGNcone{}, respectively.\footnote{The same result holds even if $L_{\rm lim, [CII]}=10^{7}~L_{\odot}$ ($L_{\rm lim, UV}=10^{9}~L_{\odot}$) - a luminosity threshold reasonably achievable with current facilities - where the median [CII] (UV) luminosities of the satellites are $3.6 \times 10^{6}$ and $7.8\times 10^{6}~L_{\odot}$ ($4.2 \times 10^{9}$ and $1.8\times 10^{10}~L_{\odot}$) in \noAGN{} and \AGNcone{}, respectively.}
For a comparison with UV data we refer the reader to \citet[][]{Di_Mascia_et_al_2021a}, where radiative transfer calculations are fully accounted for. 
Here, we focus on the comparison between our predictions and currently available ALMA data of $z\sim 6$ quasars. 

In particular, we consider the results of a recent high-resolution ALMA survey of 27 (previously [CII]-detected) quasars at $z\sim6$ by \citetalias{Venemans_et_al_2020} see also \citep[see also][]{Decarli_et_al_2017, Decarli_et_al_2018}.
The authors detected 17 companions\footnote{This number refers to satellites observed at $\Delta v \leq 1000$~km~s$^{-1}$ from the central quasar. The same number increases up to 19 for $\Delta v \leq 2000$~km~s$^{-1}$.} with $L_{\rm [CII]}\gsim10^{8} L_{\odot}$, corresponding to an average of $0.6$ companions per field.
We further notice that some of the \citetalias{Venemans_et_al_2020} observed quasars present multiple (2-3) companion galaxies.

\begin{figure*}
    \centering
    \includegraphics[width=\textwidth]{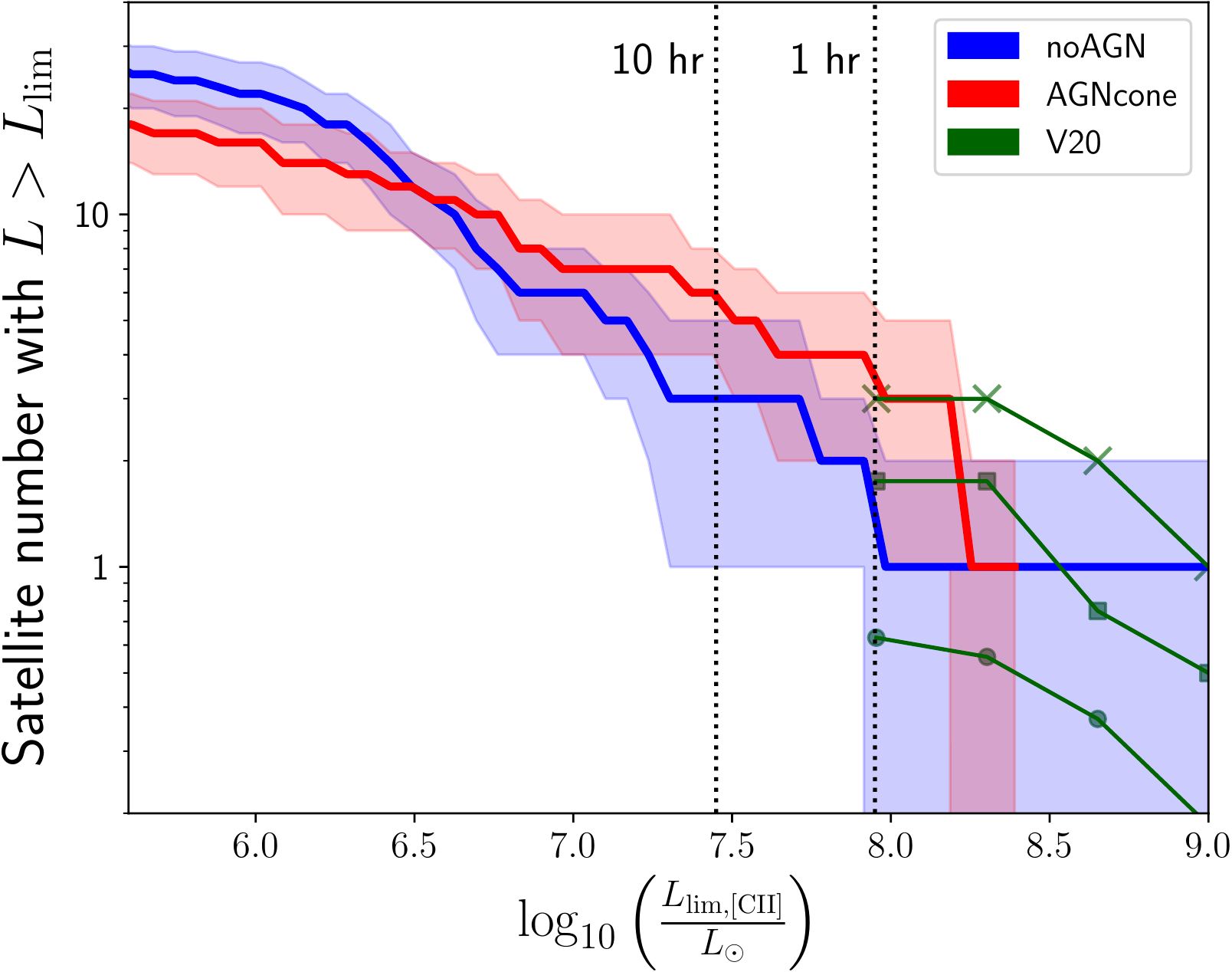}
    \caption{Number of satellites in \AGNcone{} (red) and \noAGN{} (blue) with $L_{\rm [CII]}>L_{\rm lim}$, with respect to $L_{\rm lim}$, in a volume of about $680^{3}$~kpc$^{3}$.
    Shaded areas show the related Poissonian errors \citep[with a 68~percent confidence level; ][]{Gehrels_1986}.
    From left to right, the vertical dotted lines mark the [CII] sensitivity of an ALMA observing program of 10 and 1 hour on source, respectively.
    Green symbols show the latest observational data from \citetalias{Venemans_et_al_2020}: filled circles refer to the whole sample of 17 satellites around 27 quasars, filled squares refer only to the quasar population with $L_{\rm FIR} \gtrsim 10^{13}L_{\odot}$, and Xs mark the densities around the most luminous quasar in FIR, i.e. J0305-3150.}
    \label{fig:satnum_vs_thre}
\end{figure*}

\begin{figure*}
    \centering
    \includegraphics[width=\textwidth]{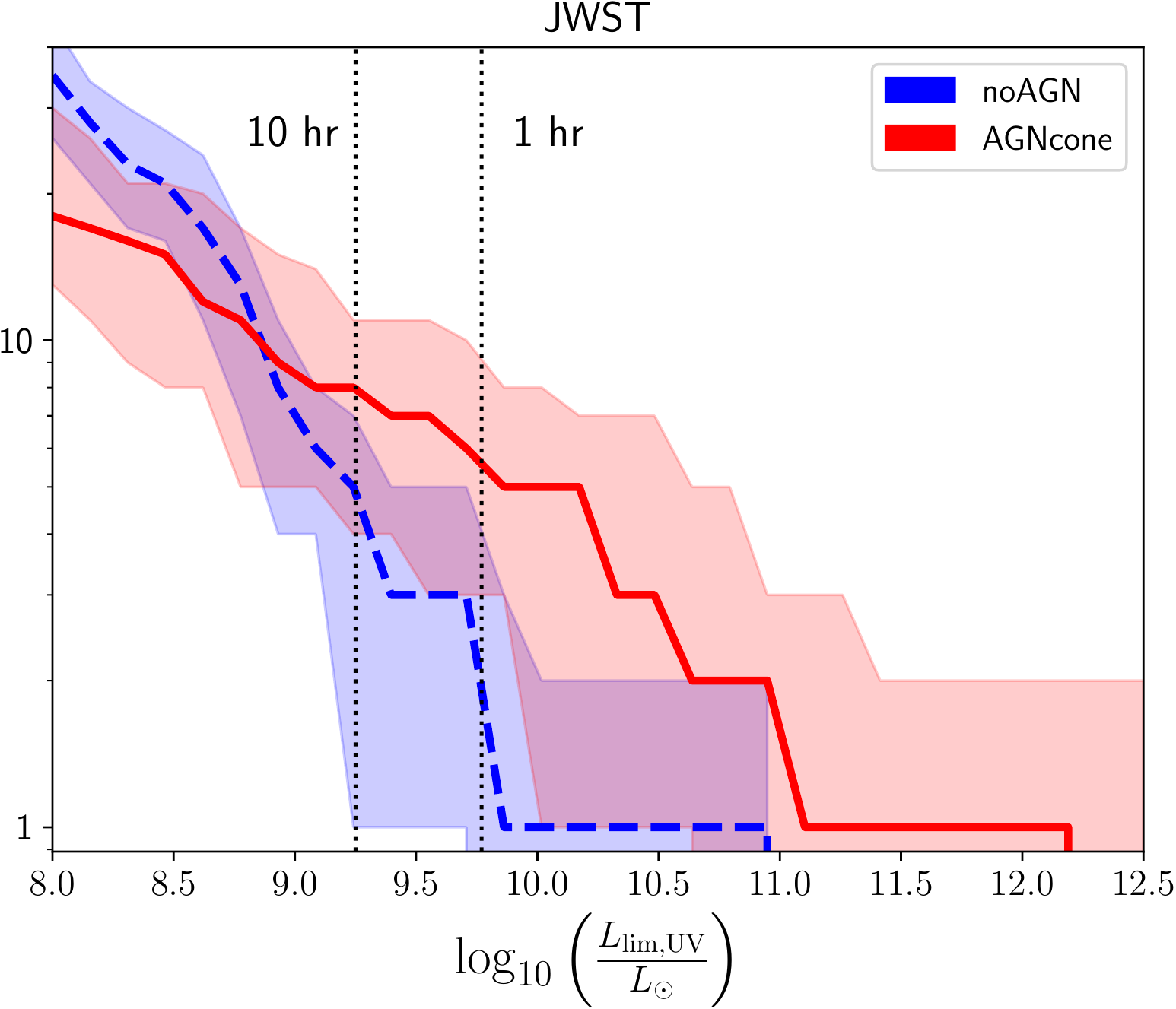}
    \caption{Analogously to Figure~\ref{fig:satnum_vs_thre}, we show the number of observable satellite in the UV band with respect to the luminosity threshold $L_{\rm lim, UV}$.
    Solid lines refer to the same UV luminosities, corrected for dust extinction of Figure~\ref{fig:observative_vs_dist}.
    The uncertainties, shown by the shaded areas, are estimated by summing the contribution of the Poissonian noise (evaluated as in Figure~\ref{fig:satnum_vs_thre}) and the variation range from the minimum to the maximum optical depth considered.
    From left to right the dotted vertical lines show the sensitivity thresholds that JWST can reach with 10~hr and 1~hr of observing time, respectively.}
    \label{fig:JWST}
\end{figure*}

In Figure~\ref{fig:satnum_vs_thre} we show the number of detectable satellites in our simulations as a function of $L_{\lim}$.
We select all those satellites above the luminosity threshold $L_{\rm lim}$ and within a spherical volume of about $680^{3}$~kpc$^{3}$, equivalent to the average volume\footnote{The companions in \citetalias{Venemans_et_al_2020} are observed in a field of view of $\pi95^{2}$~kpc$^{2}$ on the sky plane, i.e, where the sensitivity of the primary beam equals $0.2$ times the peak. 
We consider here the companions observed within $\pm 1000$~km~s$^{-1}$ from the quasar, corresponding to about $2.8$~Mpc at $z\sim6$.} observed by \citetalias{Venemans_et_al_2020}.
Three subsets of the \citetalias{Venemans_et_al_2020} companion list are also shown: the mean satellite number of the whole 27 quasar sample (green circles), the mean satellite number of the 4 most FIR luminous quasars ($L_{\rm FIR} \gtrsim 10^{13}~L_{\odot}$; green squares), and the satellites of the most FIR-luminous quasar, (J0305-3150, with $L_{\rm FIR} = 1.2\times10^{13}~L_{\odot}$; green crosses). Figure~\ref{fig:satnum_vs_thre} shows that the number of detected satellites in \citetalias{Venemans_et_al_2020} increases with the FIR luminosity of the central quasar.
A possible explanation for this trend is that more FIR-luminous quasars are likely hosted by galaxies with higher SFR, and therefore in more massive halos.
As a result, they tend to be more biased.

From a comparison between our results and the most FIR-luminous quasar in the ALMA sample, we note that \AGNcone{} \citep[where the cD has $L_{\rm FIR} \sim 5\times 10^{13}L_{\odot}$;][]{Di_Mascia_et_al_2021a} agrees quite well with ALMA data, whereas \noAGN{} predicts a number of satellites lower than observed.
This implies that, for the most luminous source in FIR, the positive feedback of quasar outflows on galaxy companions is required to reproduce observational data, even though more simulations are required to verify this effect with better statistics.

In contrast, although the average observed number count is well within the Poissonian noise (shaded areas), it is systematically lower than our predictions from both our runs.
This can be explained in two ways: (i) not all the quasars, but only the most FIR-luminous ones, live in over-dense regions; (ii) the number of satellites detected by \citetalias{Venemans_et_al_2020} only provides a lower limit to the actual number. The latter hypothesis can be related to observational artifacts. 

First, of all, the finite angular resolution of real data can prevent us from correctly quantifying the actual number of satellites around the quasar. In fact, [CII] emission in the bright, central regions of $z\sim 6$ quasars has typical sizes of about $1-5$~kpc and may therefore arise from multiple, unresolved sources. While counting satellites from simulations, we are instead assuming that all of them, even the closest to the central source (within $\sim1$~kpc), are spatially resolved. The inclusion of this observational artifact in our simulations would therefore automatically improve the agreement with observations.

We further note that, while ALMA observations probe distances up to $2-3$~Mpc from the central quasar, the high-resolution volume of our simulations is only limited to the inner $200-300$~kpc. 
If we consider ALMA observations only in a region from the center within $\pm 200$~km~s$^{-1}$ (corresponding to about $280$~kpc at $z=6$), the average number of satellites is $0.3$ (i.e., 9 galaxies around 27 quasars), resulting\footnote{These are the objects in a volume $V_{\rm A}=\pi R_{fov}^2 L \approx (200)^{3}$~kpc$^{3}$, where $R_{fov} = 95$~kpc is the radius of the ALMA field of view (fov) and $L = 280$~kpc is the displacement along the line of sight.} into a mean density of about $4 \times 10^{-8}$ satellites per kpc$^{3}$. In the noAGN and AGNcone simulations we retrieve mean densities of $2\times 10^{-8}$ satellites per kpc$^{3}$ and $6\times 10^{-8}$, respectively. This suggests that ALMA observations in this restricted volume are in remarkable agreement with the mean density expected from of our simulations.

Finally, the large asymmetry of the volume probed by ALMA ($\lesssim 100$~kpc on the sky plane versus $\gtrsim 2-3$~Mpc along the line of sight), could introduce a further bias. In some quasars observed by \citetalias{Venemans_et_al_2020} the number of satellites is larger than the average (e.g. J0305-3150 in Figure~\ref{fig:satnum_vs_thre}). 
We can explain these cases as ``fortunate sources'', where the distribution of satellites is somewhat aligned with the ALMA volume. In these sources, ALMA observations are able to detect the real and total amount of companions, which is also perfectly compatible with the number count predicted in \AGNcone{}, where the density is computed without assuming a preferred line of sight.
A more quantitative-statistical approach on this hypothesis will be the focus of a future work.

\subsection{Predictions for ALMA}
\label{subsec:ALMA}

Most of the ALMA data available so far to study $z\sim 6$ quasars are limited to shallow observations ($\lesssim 1$~hr per source). It is therefore interesting to investigate what we can learn from deeper observations. Figure~\ref{fig:satnum_vs_thre} shows that, if quasars are hosted in over-dense environments (as predicted by our simulations, by construction), we would be able to detect up to 6 satellites (10 as an upper limit) in a luminous quasar neighbour with $10$~hours of ALMA observing time. 

In the case of much deeper observing programs ($L_{\rm lim} <10^{6.5}~L_{\odot}$), the number of observable satellites in \AGNcone{} is smaller than in \noAGN{}, contrarily to what occurs in the case of shallower observations.
This inverted trend is possibly due to the two-fold effect of quasar feedback on the surrounding satellites: enhancing the SFR (and therefore the luminosity) of luminous satellites and lowering their intrinsic number.

Although such deep observations are beyond the capabilities of current (sub-)millimeter observatories, the discussed trend still implies that quasar feedback may leave signatures on the slope of the satellites number count, thus suggesting further investigations on this topic. 
However, we remark that the limited resolution of our simulation could play a role on this effect, since low mass galaxies could be differently impacted by quasar feedback in simulations with higher resolution, leading to different results.

\subsection{Predictions for JWST}

Finally, we focus on JWST in order to understand how this mission will improve our knowledge of high-$z$ quasar properties and their environment. 
Figure~\ref{fig:JWST} shows the expected number of satellites emitting in UV as a function of their luminosity $L_{\rm UV}>L_{\rm lim, UV}$.
The UV emission of satellites is corrected for dust attenuation with a dust-to-metal ratio $f_{d} = 0.08$, according to the results of \citet[][see Appendix~\ref{sec:observational_properties} for further details]{Di_Mascia_et_al_2021b}, and the shaded area takes into account both the minimum and maximum optical depth resulting from radiative transfer calculations, and the Poissonian uncertainty.

According to \AGNcone{}, we would be able to detect between 3 and 8 satellites with 1~hr of observing time, and between 4 and 10 satellites via a 10~hr observing program. 
We also report our predictions for the \noAGN{} case. 
Although we demonstrated that the absence of AGN feedback would decrease our capability of detecting quasar companions, this represents a lower limit ($1-2$ satellites with $1$~hr of observing time and $2-6$ objects with $10$~hr) for a deep JWST observation program on a single quasar.

These calculations, along with the ones presented in Figure \ref{fig:satnum_vs_thre}, demonstrate that both ALMA and JWST will be essential to improve our understanding of high-$z$ quasars environment, by increasing the number of observed satellites, probing different scales and emission processes. 
In particular, the synergy between these two observatories will be of outmost importance, since each instrument will compensate the limitation of the other one: e.g., on the one hand, the larger field of view of JWST with respect to ALMA ($\sim$ $10$~arcmin$^2$ versus $1$~arcmin$^2$) will allow us to probe broader regions around quasars, on the other hand, ALMA is able to reveal even those satellites that are dust-obscured and may elude JWST observations.


\section{Summary and conclusions}
\label{sec:conclusions}

We have investigated the effects of quasar outflows (i.e. feedback) on the visibility of companion galaxies by comparing two cosmological zoom-in simulations of a $z\sim6$ quasar, in which AGN feedback is either included or turned-off.
We have identified satellites in both runs and determined their key physical properties such as gas and stellar mass, star formation star formation rate, and metallicity.
\noindent Our findings can be summarized as follows.
\begin{itemize}
    \item Within the virial radius of the central galaxy, the number of satellites in the two runs increases with time in a very similar way, resulting in 10 companions at $z \approx 6$.
    At larger distances, the number of satellites in \AGNcone{} is 30 percent smaller than in \noAGN{};
    we suggest that this effect is due to AGN-driven outflows that, by dispersing stars and gas in the surrounding region, boost the dynamical friction and, consequently, increases the galaxy merger rate.
    The effect is negligible at smaller distances because the orbital dynamics and the CGM density are fully dominated by the presence of the central galaxy.
    \item The redshift evolution of the satellite median properties does not show striking differences between the runs.
    $M_{\rm gas}$ decreases from $10^9~\msun$ at $z\sim 10$ to $10^8~\msun$ at $z\sim 6$; 
    the SFR decreases from 1 to $0.1~\msun$~yr$^{-1}$;
    $M_*$ does not evolve appreciably from $\sim 3\times 10^7~\msun$;
    $Z$ increases from $0.05$ to $0.2$ $\rm Z_{\odot}$.
    To get a deeper insight into the effect of AGN feedback on its environment, we have thus considered a sub-sample of satellites, located at different distances and positions with respect to the center of the group and followed the redshift evolution of their intrinsic properties ($M_*$ and SFR) in both runs. 
    For all these satellites, we found that both $M_*$ and SFR grow faster in \AGNcone{} with respect to \noAGN{}.
    We argue that the SFR enhancement in those satellites engulfed by the quasar outflow can be due to an increase of the fuel available for the SF process and/or a boost in the SF efficiency due to the induced shocks. Both possibilities are linked to the kind of feedback and SF recipes implemented in the simulation: gas particles are moved away from the galaxies hosting the most accreting MBHs towards the surrounding satellites.
    \item We have developed a semi-analytical model based on momentum conservation to quantify the total energy deposited on a target galaxy by the surrounding active AGN, taking into account their duty cycles and accretion rates, the outflow orientation, the circum-galactic medium (CGM) density, and their relative distance from the target galaxy.
    We found positive feedback in numerous satellites, proportionally to the total energy received from the accreting MBHs in their whole evolutionary history.
    \item We have computed the [CII]158$\mu$m emission of satellites and studied the effect of quasar feedback on their observed number count. When compared with the most FIR-luminous quasar of the ALMA sample, the \noAGN{} run predicts a number of satellites lower than observed, while the \AGNcone{} run agrees quite well with ALMA data. This implies that, for the most FIR-luminous source, the positive feedback of quasar outflows on the SFR of galaxy companions is instrumental in reproducing observational data. However, the average number of satellites observed in the whole quasar sample is lower than our predictions for both the runs. This can be explained in two ways: (i) not all the quasars, but only the most FIR-luminous ones live in over-dense regions; (ii) the number of satellites reported by \citet{Venemans_et_al_2020} only provides a lower limit to the actual number, as a consequence of observational artifacts (e.g. finite angular resolution and anisotropic volume probed by ALMA).
    \item We predict that JWST will double the current ALMA detections after just $1$~hr of observing time and will allow us to directly observe a major part of the quasar neighbour, avoiding any bias introduced by the small ALMA field of view. Still, ALMA will be necessary to reveal those satellites around $z\sim 6$ quasars that are dust-obscured and may elude JWST observations. 
    We find that with $10$~hr of observing time, ALMA and JWST will detect up to $10$ satellites per sources.
\end{itemize}

In conclusion, we did not find any evidence of AGN-driven quenching on the star formation of satellites surrounding high-$z$ quasars. Thus, we rule out a scenario in which the detection of companions can be undermined by the external feedback.
AGN feedback could be even instrumental to explain the high satellite number count observed around the most FIR luminous quasars and -- after correcting data for ALMA observational biases -- in the whole population observed so far.


\section*{acknowledgements}
The authors thank Bram Venemans and Roberto Decarli for helpful insights on ALMA [CII] data. SG acknowledges support from the ASI-INAF n. 2018-31-HH.0 grant and PRIN-MIUR 2017. AF and SC acknowledge support from the ERC Advanced Grant INTERSTELLAR H2020/740120. Partial support from the Carl Friedrich von Siemens-Forschungspreis der Alexander von Humboldt-Stiftung Research Award (AF) is kindly acknowledged.
We gratefully acknowledge computational resources of the Center for High Performance Computing (CHPC) at SNS.
AL acknowledges funding from MIUR under the grantPRIN 2017-MB8AEZ.
PB acknowledges support from the Brazilian Agency FAPESP (grants 2016/01355-5, 2016/22183-8).
The authors greatly thank the anonymous referee for useful comments which improved the quality of this manuscript.

\section*{Data Availability Statement}
The data underlying this article will be shared on reasonable request to the corresponding author.

\bibliography{bibliography.bib}


\appendix

\section{Accretion-weighted centre}
\label{sec:AccretionCentre}
Given the large number of active MBHs in the simulation, each one characterized by different positions, duty cycles, and accretion rates, the overall AGN activity in the galaxy proto-cluster is the result of a complex spatial distribution and time-dependent superposition of individual, partially overlapping, AGN activities.
We thus define, for each snapshot of the simulation \AGNcone{}, a center of reference that takes into account both the position and the accretion rate of all the AGN in the simulation. We refer to this point in the simulation volume as the ``accretion-weighted center", ${\bf c}_{aw}$, of the galaxy proto-cluster.

\begin{figure*}
    \centering
    \includegraphics[width=\textwidth]{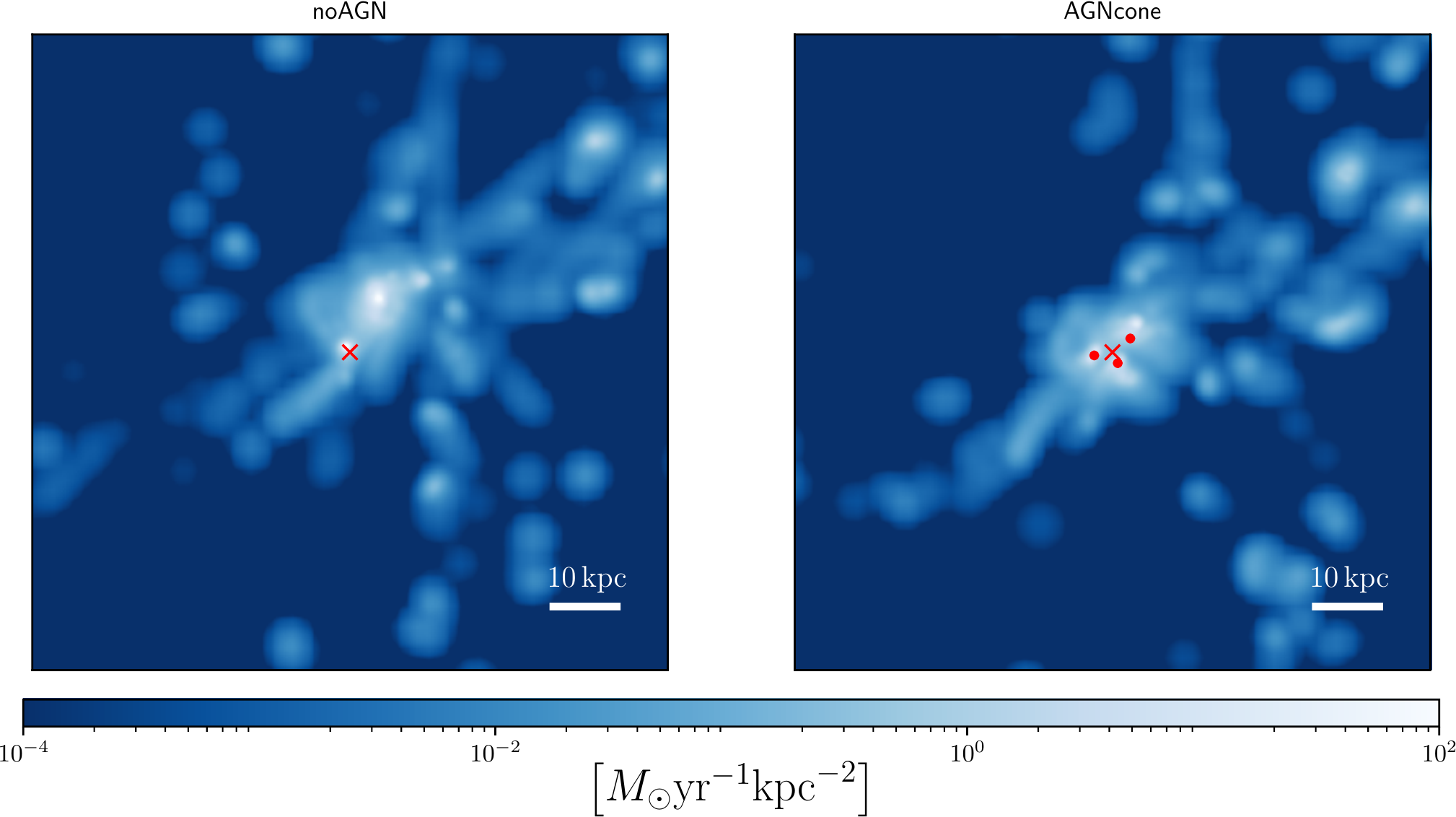}
    \caption{SFR density map of the central $100$~kpc at $z=6.3$ for \noAGN{} ({\it left panel}) and \AGNcone{} ({\it right panel}).
    The accretion-weighted mean centre is marked in each panel with a red cross, along with the position of the most accreting MBHs ($\dot{M}_{\rm acc}>1~ \msun$~yr$^{-1}$).}
    \label{fig:bhmap}
\end{figure*}

Despite the numerous MBHs in the simulation volume, their luminosity remains almost irrelevant for most of the time.
In particular, at $z<10$, beyond the four most accreting MBHs, rarely $\dot{M}_{\rm acc}>0.1~\msun$~yr$^{-1}$.
However, in order to be more conservative, we compute ${\bf c}_{aw}$, taking into account the eight most accreting MBHs in the simulation volume, weighted for the accretion rate of each considered MBH:

\begin{equation}
    {\bf c}_{aw}(t) = \frac{\sum_{i=1}^8 \dot{M}_{{\rm acc}, i}(t-\Delta t_{\rm snap}){\bf r}_{i}(t-\Delta t_{\rm snap})}{\sum_{i=1}^8 \dot{M}_{{\rm acc}, i}(t-\Delta t_{\rm snap})},
    \label{eq:caw}
\end{equation}
where $\Delta t_{\rm snap} \simeq 2 \times 10^{7}$~yr is the minimum time spacing between two snapshots in the run; in other words, to compute ${\bf c}_{aw}$ at a given snapshot (i.e. at a given time $t$), we consider the accretion rate and the MBH positions in the previous snapshot (i.e. at the time $t-\Delta t_{\rm snap}$).
This method allows us to take into account both the time delay between the launch of the outflow and the arrival of the outflowing gas on the satellites, and the time necessary to develop the possible feedback-driven effects on the satellite properties (e.g. $M_{*}$ and SFR).
Indeed, if we consider, for a typical galaxy in this study, a scale distance from the feedback source of about $100$~kpc and an initial wind speed of $10^{4}$~km s~$^{-1}$, the time scale needed by the outflow to reach the target galaxy is $\Delta t_{\rm wind} > 10^{7}$~yr $\sim\Delta t_{\rm snap}$.
Moreover, the typical time-sale needed to convert half of the dense gas mass into stars is $\Delta t_{\rm SF}\sim10^{7}-10^{8}$~yr $\sim \Delta t_{\rm snap}$ \citep[see, e.g.,][]{Gao_Solomon_2004, Wu_et_al_2005, Hartmann_2009}.

The determination of ${\bf c}_{aw}$ occurs in two steps: at first ($i$) the eight most accreting MBHs are found in each snapshot and then ($ii$) their location is followed in the next snapshot by selecting either the position of the MBH with the same id, if the MBH is still present, or the position of the descendent MBH, if a coalescence event occurred in the time interval between the two consecutive snapshots.

To find a proper counterpart of ${\bf c}_{aw}$ in the \noAGN{} run, where we clearly cannot apply the method described above, we proceed as follows:
($i$) we consider the positions of the MBHs that contributes to ${\bf c}_{aw}$;
($ii$) we assign a DM halo to each MBH, only on the basis of its position;
($iii$) we identify the corresponding halo in the \noAGN{} run by cross-matching the DM particle IDs in the two runs, selecting the halo which shares the largest fraction of particles;
($iv$) in this run, we choose the halo that maximizes the ratio $N_{\rm s}/N_{\rm DM}$, where $N_{\rm s}$ is the number of particles shared among the halos and $N_{\rm DM}$ is the total number of DM particles in the halo of \noAGN{} run;
($v$) we save the centre of mass of the selected halo in \noAGN{}, as the equivalent position of the MBH in \AGNcone{};
($vi$) the position of ${\bf c}_{aw}$ is finally determined through Equation~\ref{eq:caw}, by setting $\dot{M}_{{\rm acc}, i}$ to the corresponding value for the $i$-th MBH in \AGNcone{}. 

Figure~\ref{fig:bhmap} shows the location of the ${\bf c}_{aw}$ at $z=6.3$, over-imposed on the SFR density map, and marked with a red cross; the most accreting MBHs are instead denoted by red filled circles.

\section{Observational properties}
\label{sec:observational_properties}

In this Appendix, we describe the modelling adopted to compute the [CII] and UV luminosities of $z\sim 6$ quasar companions. 

\subsection{[CII] luminosity}

The expected [CII] luminosity for the satellite sample is evaluated through the relation
\begin{equation}
    \log_{10}{(L_{[CII]}/L_\odot)} = 7.0 + 1.2s + 0.021t + 0.012st - 0.74 t^2,
    \label{eq:vallini}
\end{equation}
where $s = \log_{10}{(\rm SFR/\msun yr^{-1})}$ and $t=\log_{10}{(Z/Z_\odot)}$. This relation, provided by \citet{Yue_et_al_2015}, is based on the \citet{Vallini_et_al_2013, Vallini_et_al_2015} model that has been tested against data of $z\sim 6$ galaxies and well reproduces the [CII]-SFR relation at high-$z$ \citep[][]{Pallottini_et_al_2017, Pallottini_et_al_2019}. However, it has the limit of not considering the effect of X-ray on the [CII] emission \citep[e.g.][]{Meijerink_et_al_2007, Langer_Pineda_2015}.

\subsection{UV luminosity}

We compute the satellites expected UV luminosity at $1450$~\AA~by considering star emission only in \noAGN{}, and both stellar and AGN contributions in \AGNcone{}, i.e. $L_{\rm UV}=L_{\rm UV}^{\rm star}+L_{\rm UV}^{\rm AGN}$, where
\begin{equation}
    L_{\rm UV}^{\rm star} = 2.2\times 10^{43}~SFR,
    \label{eq:SFUV}
\end{equation}
according to \citet{Kennicutt_Evans_2012}, and 
\begin{equation}
L_{\rm UV}^{AGN}=f_{\rm UV}~L_{\rm bol}.
\label{eq:UVluminosity2}
\end{equation}
$f_{\rm UV}$ is the bolometric corrections provided by \citet{Shen_et_al_2020},
$L_{\rm bol} = \epsilon_{r}(1-\epsilon_{f})\dot{M}_{\rm acc}c^{2}$ is the bolometric luminosity, and $\dot{M}_{\rm BH}$ is the sum of the accretion rates of every MBHs within $\beta r_{\rm vir}$.

After this, we consider dust attenuation by applying an extinction coefficient $e^{-\tau_{\rm eff}}$ to the UV emission.
In particular, we consider the minimum and maximum values of the optical depth, as they are evaluated by \citet{Di_Mascia_et_al_2021a} on the same \AGNcone{} and \noAGN{} runs, through radiative transfer calculations, along 6 different lines of sight.
We then average these values ($\tau = 1.8 - 2$ for \noAGN{} and $\tau = 1.5 - 2.7$ for \AGNcone{}, with a dust-to-metal ratio $f_{d} = 0.08$; \citealt[][]{Behrens_et_al_2018}) and apply the resulting $\tau_{\rm eff}$ to each satellite.

\end{document}